\documentclass[jcp,twocolumn,amsmath,amsfonts,amssymb,showpacs,final]{revtex4-1}
\usepackage{graphicx,graphicx,epsfig}

\usepackage{graphicx}
\usepackage{bm}
\usepackage[all]{xy}

\newcommand{\qed}{\nobreak \ifvmode \relax \else
      \ifdim\lastskip<1.5em \hskip-\lastskip
      \hskip1.5em plus0em minus0.5em \fi \nobreak
      \vrule height0.75em width0.5em depth0.25em\fi}
\newcommand{\beq}{{\begin{equation}}}
\newcommand{\eeq}{{\end{equation}}}

\begin{document}
\title{Quantization and Fractional Quantization of Currents in
  Periodically Driven Stochastic Systems II: Full Counting Statistics}

\author{Vladimir~Y.~Chernyak}
\affiliation{Department of Chemistry, Wayne State University, 5101 Cass Avenue, Detroit, MI 48202,
Department of Mathematics, Wayne State University, 656 W. Kirby, Detroit, MI 48202,
and Theoretical Division, Los Alamos National Laboratory, Los Alamos, NM 87545 USA}
\author{John~R.~Klein}
\affiliation{Department of Mathematics, Wayne State University, 656 W. Kirby, Detroit, MI 48202,
Center for Nonlinear Studies, Los Alamos National Laboratory, Los Alamos, NM 87545 USA, and New Mexico Consortium, Los Alamos, NM 87545 USA}
\author{Nikolai~A.~Sinitsyn}
\affiliation{Theoretical Division, Los Alamos National Laboratory, Los Alamos, NM 87545 USA and New Mexico Consortium, Los Alamos, NM 87545 USA}


\begin{abstract}
We study  Markovian stochastic motion on a graph with finite number of nodes and adiabatically periodically driven transition rates. We show that,  under general conditions, the quantized currents that appear at low temperatures are a manifestation of topological invariants in the counting statistics of currents. This observation provides an approach for classification of topological properties of the counting statistics, as well as for extensions of the phenomenon of the robust quantization of currents at low temperatures to the properties of the counting statistics which persist to finite temperatures.
\end{abstract}

\date{\today}

\maketitle

\section{Introduction}
\label{intro-II}

This is the second manuscript of the series devoted to quantization of currents in periodically driven stochastic systems. In the previous article \cite{first}, the main object of our study was the average current generated per period $\tau_{D}$ of the driving protocol. The average current, however, is only one of the characteristics of the observable stochastic currents in mesoscopic systems. The current, ${\bm Q}$, generated by a periodic driving protocol, can be treated as a  random variable.  In realistic experiments, mesoscopic stochastic fluctuations of currents (fluxes) can be observed via accumulation of sufficient statistics so that one can study not only the mean and the variance but also the values of a large number of higher cumulants, associated with the current distribution.

The theory of the full counting statistics of currents in periodically driven stochastic systems has  attracted lot of attention \cite{ sukhorukov-07Nat,nazarov-03,sukhorukov-04,ft-exp,ganeshan-11,cao1,cao2,cao3,sinitsyn-07epl,ohkubo-08jcp,ohkubo-08jstat,sinitsyn-09review,ohkubo-10jpa,sinitsyn-10jstat,sinitsyn-11pre, buttiker-1,buttiker-2,buttiker-3,abanov-10epl,ren-10prl,sinitsyn-guc}. This interest was motivated by experiments with molecular motors \cite{Leigh-03,turnstile,hanggi-09rev} and mesoscopic electric circuits \cite{sukhorukov-07Nat,nazarov-03,sukhorukov-04,ft-exp}, as well as applications in  soft matter \cite{cao1,cao2,cao3,sinitsyn-09pnas,szabo-06jcp,astumian-11rev}, nonequilibrium thermodynamics \cite{sinitsyn-11pre,clausius,andrieux,saito-exp} and life sciences \cite{kamenev-06,sinitsyn-guc}. Therefore, the central object of study in the present article is not the average current, but rather the complete {\it  counting statistics} of currents, which is represented by the
probability distribution function ${\cal P}({\bm Q})$ of currents ${\bm Q}$. Specifically, we will concentrate on the
 {\it Large Deviation} (LD) function, ${\cal S}({\bm Q})$, that describes the long-time  asymptotic behavior of the probability distribution function, given by \cite{ldf,vankampen}
 \begin{eqnarray}
\label{P-LD-form} {\cal P} ({\bm Q})\sim e^{-N{\cal S}({\bm Q})},
\end{eqnarray}
where $N$ is the number of repeated full cycles of the driving protocol. This function is independent of initial conditions. The universality property, given by Eq.~(\ref{P-LD-form}), is known to become exact in the ``thermodynamic limit", $N \rightarrow \infty$.

In this manuscript we will explore the property of the counting statistics that, to our knowledge, has been never considered before in the framework of classical stochastic Markov chain models. We will show that certain topological invariants, in our case the Chern classes, can be associated with counting statistics of currents in periodically driven systems. We are only aware of similar observations in quantum mechanical systems at zero temperature \cite{kamenev,makhlin-mirlin}, which originate from essentially different initial physical assumptions.

The first goal of the present manuscript is to explore simple models, in which the full counting statistics can be obtained explicitly and the corresponding topological properties can be clearly illustrated. Our second goal is to demonstrate that integer quantization of currents is a special manifestation of topological effects in the full counting statistics. This relation is important since in the first manuscript of the series \cite{first}, we classified integer quantized currents. By relating quantized average currents to topological invariants of the full counting statistics we build an approach for classification of such invariants.

The manuscript is organized as follows. In section~\ref{generating-twisted}, we introduce the basic notation, including the definition of the generating function and its evolution equation. In section~\ref{toys}, we consider two simple models, one with integer, and one with fractional average current quantization. We will study in some detail the low-temperature limit for these models. In particular, we will obtain the generating functions associated with the current distributions and demonstrate the topological nature of certain ingredients in the expressions by relating them to Chern classes. In section~\ref{generic-networks} we extend the picture of the integer average current quantization to the level of the LD function ${\cal S}({\bm Q})$. This will be done for an arbitrary network, whose geometry is described by a general type of a finite connected non-oriented graph. We will show that under the condition of the average current integer quantization, the expression for the generating function (for an arbitrary graph geometry) involves a set ${\bm n}({\bm s})=(n_{\alpha}({\bm s})|\alpha\in X_{1})$ of integer-valued topological invariants, identified as the Chern numbers. The main result of this section is represented by the claim that the LD function in the low temperature adiabatic limit ${\cal S}({\bm Q})={\cal S}^{A}({\bm Q}-{\bm n}({\bm s}))$ is determined by an even function ${\cal S}^{A}({\bm Q})$, with an integer-valued shift, determined by a topological invariant ${\bm n}({\bm s})$, identified as a set of Chern numbers. This result not only provides a Chern-class interpretation of the average current quantization, but also establishes a stronger property of the current PDF, i.e., that the average current is the only odd cumulant that survives in the low-temperature adiabatic limit. In section~\ref{chern-II}, we demonstrate how the robust parameter subspace ${\cal M}_{X}^{r}$, defined in the previous article, arises naturally when the LD function is studied in terms of the corresponding generating function, with the focus on its global properties. We will then develop topological interpretation of the (fractionally) quantized currents. The conclusion section summarizes our findings and discusses possible future research directions. Two appendixes serve the goal of providing some very basic information on Chern classes, which is an attempt to make the manuscript self-contained.

\section{Generating functions and twisted master operators}
\label{generating-twisted}

In this manuscript we show that integer quantization of pumped currents in the low-temperature adiabatic limit is a consequence of a more general property of the long-time asymptotics of the current probability distribution ${\cal P}({\bm Q})$.

We start with a case of general (not necessarily periodic) pumping. The easiest way to study the current probability distribution is to introduce the corresponding generating function
\begin{eqnarray}
\label{define-Z} Z({\bm\lambda};t) &=& \sum_{{\bm q}}{\cal P}({\bm q};t)
\prod_{\alpha\in X_{1}} \lambda_{\alpha}^{q_{\alpha}} \nonumber \\ &=&\sum_{{\bm q}}{\cal P}({\bm q};t)
e^{i\sum_{\alpha\in X_{1}}\chi_{\alpha}{q_{\alpha}}},
\end{eqnarray}
where ${\bm\chi}=(\chi_{\alpha}|\alpha\in X_{1})$ is a
multi-variant argument of the generating function in  additive form, represented by a
vector spanned on the graph links. We denote by ${\bm\lambda}=(\lambda_{\alpha}|\alpha\in X_{1})$ its multiplicative counterpart; they are naturally related by ${\bm\lambda}=e^{i{\bm \chi}}$. The space in which argument ${\bm\lambda}$ resides is a multidimensional torus (a cartesian product over the graph links $\alpha$ of the circles $|\lambda_{\alpha}|=1$) is denoted by ${\cal T}$. When a link $\alpha$ with $\partial\alpha=\{i,j\}$ is not a self link, i.e., it connects two different nodes $i$ and $j$, it is convenient to represent $\chi_{\alpha}$ by $\chi_{i\alpha}$ or $\chi_{j\alpha}$, with $\chi_{i\alpha} =-\chi_{j\alpha}$. For the multiplicative counterpart we use the notation $\lambda_{i\alpha}$ or $\lambda_{j\alpha}$, with $\lambda_{i\alpha}\lambda_{j\alpha}=1$. The generating function $Z$ is a periodic function with respect to parameters $\chi_{j\alpha}$ with period $2\pi$. Stated equivalently, $Z$ is well-defined (i.e., single-valued) function of ${\bm\lambda}$. These properties reflect the discrete nature of the transition events through the links.

Similar to the case of the average current, the generating function has a useful operator representation
\begin{eqnarray}
\label{Z-operator} Z({\bm\lambda};t)=\sum_{k,j\in X_{0}}\left({\rm T}\exp\left(\int_{0}^{t}dt'\hat{H}_{\bm\lambda}(t')\right)\right)_{kj} \rho_{j}(0),
\end{eqnarray}
where the {\it twisted master operator} $\hat{H}_{{\bm\lambda}}({\bm x})$ has the form \cite{derrida}
\begin{eqnarray}
\label{H-twisted} (\hat{H}_{{\bm\lambda}}{\bm\rho})_{i}&=&\sum_{j\ne i}\sum_{\alpha\in X_{1}}^{\partial\alpha=\{i,j\}}e^{-\beta W_{\alpha}}\left(e^{\beta E_{j}}\lambda_{j\alpha}\rho_{j}-e^{\beta E_{i}}\rho_{i}\right) \nonumber \\ &+& \sum_{\alpha\in X_{1}}^{\partial\alpha=\{i\}}\left(\lambda_{\alpha}+ \lambda_{\alpha}^{-1}-2\right)e^{\beta(E_i-W_{\alpha})}\rho_{i}.
\end{eqnarray}
The word ``twisted'' here reflects the way the operator $\hat{H}_{{\bm\lambda}}$ is constructed, namely
it  is obtained from the master operator by multiplying (twisting) the rates $k_{j\alpha}$ in its off-diagonal components, responsible for the transitions through the links $\alpha$, with the factors $\lambda_{j\alpha}$. The diagonal elements of the twisted master operator $\hat{H}_{{\bm\lambda}}$ are the same as in $\hat{H}$, i.e., twisting affects the off-diagonal components only. Twisting allows the counting of jumps through links. Note that the relations $\lambda_{i\alpha}\lambda_{j\alpha}=1$ for the links $\alpha$ with $\partial\alpha=\{i,j\}$, i.e. connecting the nodes $i$ and $j$, results in counting the fluxes, i.e. the differences of the numbers of jumps along the links in two different directions. Also note that relaxing the above conditions would result in a more detailed generating functions that allows counting the number of jumps along any link in each direction separately. However, in this manuscript we focus on the fluxes, and hence the above relations are always in place.

Direct inspection of the matrix elements $(H_{{\bm\lambda}})_{ij}$ of the twisted master operator shows that, provided $|\lambda_{\alpha}|=1$ for all links $\alpha\in X_{1}$, or equivalently ${\bm\lambda}\in {\cal T}$, the operator $\hat{H}_{\bm\lambda}$ is Hermitian with respect to the weighted scalar product in the population vector space, defined by
\begin{eqnarray}
\label{weighted-sclar-populations} \left\langle{\bm\rho}|{\bm\rho}'\right\rangle_{{\bm\kappa}}=\sum_{j\in X_{0}}\kappa_{j}\rho_{j}^{*}\rho'_{j}=\sum_{j\in X_{0}}e^{\beta E_{j}}\rho_{j}^{*}\rho'_{j},
\end{eqnarray}
and we reiterate that, according to the notation, introduced in the first manuscript of the series $\kappa_{j}=e^{\beta E_{j}}$ and $g_{\alpha}=e^{-\beta W_{\alpha}}$, so that the rates are parametrized by $k_{j\alpha}=g_{\alpha}\kappa_{j}$. The Hermitian nature of the twisted master operators $\hat{H}_{\bm\lambda}$ is important for application of the adiabatic limit, since it implies that all eigenvalues of $\hat{H}_{\bm\lambda}$ are real so that that the notion of a ground state as an eigenvector with the lowest absolute eigenvalue is in place. It is also important to note that the untwisted master operator $\hat{H}=\hat{H}_{{\bm 1}}$ is also real, and hence its eigenmodes, i.e. the Boltzmann distribution and relaxation modes are represented by vectors with real components. The eigenmodes of a twisted operator $\hat{H}_{{\bm\lambda}}$ in the generic case ${\bm\lambda}\ne {\bm 1}$ are represented by vectors with generally complex components. It is the latter circumstance that allows the topological invariants, namely the first Chern class to participate in the game.

In the case of periodic driving, we choose the observation time, $t_{0}=N\tau_{D}$, to be an integer multiple $N$ of the protocol time period $\tau_{D}$, and note that for large $N$ the expression for the generating function [Eq.~(\ref{Z-operator})] is dominated by the eigenmode of the twisted evolution operator
\begin{equation}
\label{define-U-twisted} \hat{U}_{{\bm\lambda}}({\bm s};\tau_{D})\equiv{\rm T}\exp\left(\tau_{D}\int_{0}^{1}d\tau\hat{H}_{{\bm\lambda}}({\bm s}(\tau))\right),
\end{equation}
with the largest absolute value of the eigenvalue, the latter denoted by $z_{{\bm\lambda}}$, so that we have for the generating function in the large $N$ limit with the exponential accuracy
\begin{eqnarray}
\label{Z-LD-form} Z({\bm\lambda};N)\sim z_{{\bm\lambda}}^N=e^{N\omega_{{\bm\lambda}}}.
\end{eqnarray}
where we have introduced the {\it cumulant generating function} $\omega_{{\bm\lambda}}$ defined by $z_{{\bm\lambda}}=e^{\omega_{{\bm\lambda}}}$. The LD function, ${\cal S}({\bm Q})$, is related to $\omega_{{\bm\lambda}}$ via the inverse Legendre transform, ${\cal S}({\bm Q})={\rm max}_{{\bm\lambda}} \left(-{\bm \chi Q} + \omega_{{\bm\lambda}}\right)$. Since ${\bm\lambda}=e^{i{\bm\chi}}$,  we may  write $\omega_{\bm\chi}$ instead of $\omega_{\bm\lambda}$ with the same meaning.

The generating function $z_{{\bm\lambda}}$ possesses a gauge invariance that reflects conservation of currents at long times. A gauge transformation, naturally labeled by a vector ${\bm \varphi}=(\varphi_{j}|j\in X_{0})$ in the additive representation and ${\bm h}=(h_{j}|j\in X_{0})$, with $h_{j}=e^{i\chi_{j}}$ in its multiplicative counterpart,  transforms ${\bm\chi}$ to ${\bm\chi}'$ and ${\bm\lambda}$ to ${\bm\lambda}'$, so that for a link $\alpha$ with $\partial\alpha=\{i,j\}$, we have $\chi'_{i\alpha}=\chi_{i\alpha} - \varphi_{i} +\varphi_{j}$, and $\lambda_{i\alpha}'=h_{i}^{-1}\lambda_{i\alpha}h_{j}$, respectively. The gauge invariance
\begin{eqnarray}
\label{gauge-invariance} z_{{\bm\chi}'}=z_{{\bm\chi}}, \;\;\; z_{{\bm\lambda}'}=z_{{\bm\lambda}}
\end{eqnarray}
follows from an obvious fact
$\hat{H}_{{\bm\chi}',ij}=e^{-i\varphi_{i}}\hat{H}_{{\bm\chi},ij}e^{i\varphi_{j}}$, or equivalently $\hat{H}_{{\bm\lambda}',ij}=h_{i}^{-1}\hat{H}_{{\bm\lambda},ij}h_{j}$ which does not change the eigenvalues of the evolution operator, given by Eq.~(\ref{define-U-twisted}). This can be verified directly, and reflects the law of current conservation (or, in other words, the continuity equation)
\begin{eqnarray}
\label{current-conservation} \partial {\bm Q}=0,
\end{eqnarray}
the latter representing a discrete counterpart of the well known equation ${\rm div}{\bm Q}=0$.

\section{Toy Models}
\label{toys}

The goal of this section is to illustrate the general concepts using some simple examples. To that end we consider in detail two simple models. The first one that contains two nodes connected by two links has been considered in the first manuscript of the series to illustrate the main concepts relevant for the average current quantization. The second model extends the first one to the case of two nodes connected by an arbitrary number $N$ of links. In this section we calculate explicitly the generating function $Z({\bm\lambda};N)$ and its long-time LD counterpart $z_{{\bm\lambda}}$ [related by Eq.~(\ref{Z-LD-form})] for these simple models and demonstrate in this simple case how the average current is related to the topological invariants of the full counting statistics. The general case is considered in section~\ref{generic-networks}. Since in the physics literature it is  common to use the additive representation for the argument of the generating function, throughout this section the additive argument ${\bm\chi}$ will be used, whereas in section~\ref{generic-networks} we switch to the multiplicative argument, as more appropriate and convenient for dealing with the case of a general network. We reiterate that the multiplicative representation is related to its additive counterpart by ${\bm\lambda}=e^{i{\bm\chi}}$.

\subsection{Markov-chain on a  2-nodes-2-links graph}
\label{model2}
\begin{figure}
\centerline{\includegraphics[width=2.0in]{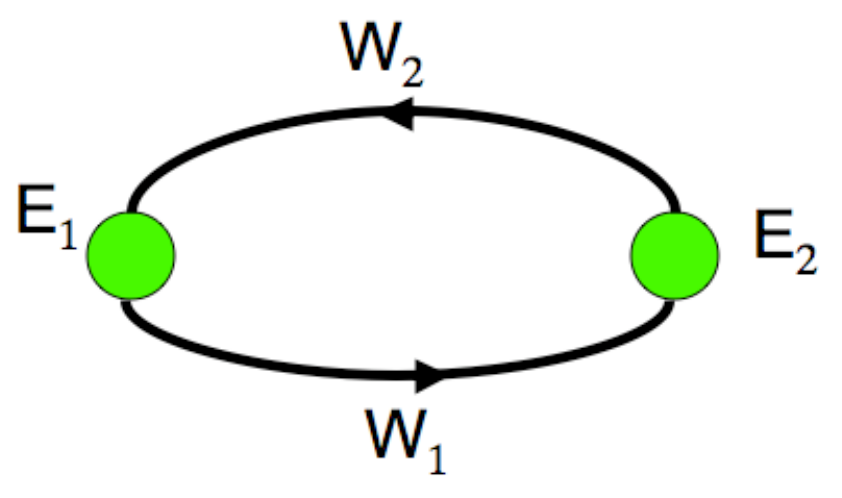}}
  \caption{The ``2-nodes-2-links'' model of stochastic transitions between two states along two possible paths. Different paths are
characterized by the values of the corresponding potential barriers $W_1$ and $W_2$. The two states are characterized by
the values of their well depths, $E_1$ and $E_2$. Periodic
modulation of these parameters leads to appearance of a nonzero, on average, circulating current in the preferred (clockwise or counterclockwise) direction. }
\label{2level-1}
\end{figure}

Our simplest model with a pair of nodes connected by a pair of links is model in Fig.~\ref{2level-1}: a particle randomly jumps between two sites along two different paths. The vector ${\bm \rho}=(\rho_1,\rho_2)$, of the node populations $\rho_1$ and $\rho_2$ evolves according to the master
equation,
\begin{equation}
\frac{d}{dt}
\left( \begin{array}{c}
\rho_1 \\
\rho_2
\end{array} \right) = \hat{H} \left( \begin{array}{c}
\rho_1 \\
\rho_2
\end{array} \right),
\label{partial}
\end{equation}
with the master operator
\begin{equation}
\hat{H}=
\left( \begin{array}{cc}
-\kappa_1(g_1+g_2) & \kappa_2(g_1+g_2) \\
\kappa_1(g_1+g_2) & -\kappa_2(g_1+g_2)
\end{array} \right),
\label{master-simple}
\end{equation}
where the kinetic rates  are expressed in terms of the node energies $E_{j}$ and barrier heights $W_{\alpha}$ by $k_{j\alpha}=\kappa_{j}g_{\alpha}$, $\kappa_j=e^{\beta E_{j}}$, $g_{\alpha}=e^{-\beta W_{\alpha}}$, with $\beta =1/(k_BT)$ being an inverse temperature.

The main subject of our studies presented in the first manuscript of the series, was the average current per period, given by
\begin{equation}
\label{currentQ-two} {\bm Q}=N^{-1}\int_0^{N\tau_D} dt {\bm J}(t), \quad N\to \infty.
\end{equation}

In the present manuscript we rather treat both ${\bm Q}$ and ${\bm J}(t)$ as vectors with fluctuating components. Note that, in the toy model, current conservation requires that $Q_{1}=Q_{2}=Q$, so it is enough to study the current component, $Q$, only through the first link.

\subsection{Generating functions and twisted master operators}
\label{gener-funct-toy}

We will study the currents that circulate in a counterclockwise direction, see Fig~\ref{2level-1}, under the action of a periodic driving of parameters. The currents are stochastic and, generally, can be characterized by probabilities,  $ \rho_{j,n}(t)$, that at time $t$ the system is at node $j$, having performed $n$ transitions during the time segment $[0,t]$ through the link $1$ (with the barrier $W_{1}$), counting counterclock/clockwise transitions with the $\pm$ sign, respectively.

The Master Equation for $\rho_{j,n}(t)$ is given by
\begin{equation}
\begin{array}{l}
\dot{\rho}_{1,n} = -(k_{11}+k_{12})\rho_{1,n} +k_{21}\rho_{2,(n-1)}+k_{22}\rho_{2,n},\\
\\
\dot{\rho}_{2,n} = -(k_{21}+k_{22})\rho_{2,n} +k_{11}\rho_{1,(n+1)}+k_{12}\rho_{1,n}.
\end{array}
\label{master-2}
\end{equation}
Recalling that $k_{j\alpha}=\kappa_j g_{\alpha}$, followed by multiplying Eq.~(\ref{master-2}) by $e^{i\chi n}$ and further performing the summation over $n$ we find

\begin{equation}
\frac{d}{dt}
\left( \begin{array}{c}
Z_1 \\
Z_2
\end{array} \right) = \hat{H}_{\chi} \left( \begin{array}{c}
Z_1 \\
Z_2
\end{array} \right),
\label{twisted-me}
\end{equation}
where
\begin{equation}
Z_j=Z_j(\chi,t) = \sum_{n=-\infty}^{\infty} \rho_{j,n}e^{in\chi},
\quad j=1,2,
\label{z12}
\end{equation}
and where
\begin{equation}
\hat{H}_\chi=\left(
\begin{array}{ll}
-\kappa_{1}(g_{1}+g_{2})& \kappa_{2}(g_{1}e^{-i\chi}+g_{2})\\
 \kappa_{1}(g_{1}e^{i\chi}+g_{2}) & -\kappa_{2}(g_{1}+g_{2})
\end{array}\right),
\label{ham-chi2}
\end{equation}
 where $\hat{H}_{\chi}$ is the twisted master operator, in accordance with its general definition, given by Eq.~(\ref{H-twisted}).

After many periods of the driving protocols the information on the initial state of the probability vector is dissipating, and the information on the currents per period of the driving protocol is contained in the largest eigenvalue $z_{\chi}$ of the evolution operator
\begin{equation}
\label{define-U-twisted1} \hat{U}_{{\chi}} \equiv {\rm {\hat{ T}}}
\exp\left(\int_{0}^{\tau_{D}}dt \hat{H}_{{\chi}}(t)\right),
\end{equation}
with the twisted master operator, given by Eq.~(\ref{ham-chi2}), so that after $N$ periods of driving, and for large $N$, the generating function of currents is given by
\begin{equation}
\label{gen-2nodes} Z(\chi;N)= Z_{1}(\chi;N)+Z_{2}(\chi;N) \sim  z_{\chi}^N,
\end{equation}
which represents a particular case of Eq.~(\ref{Z-LD-form}).

\subsection{Symmetries of the twisted master operator and counting statistics}
\label{toy-gauge-hermitian}

The closed nature of the considered stochastic process implies  that the particles are not entering and leaving the network, there is actually one particle, jumping between the graph nodes along the links, imposing current conservation in the long-time limit. The Arrhenius parametrization of the transition rates $k_{j\alpha}$ that reflects ``instantaneous'' detailed balance implies that the only non-equilibrium feature of the process that leads to the current generation is the time-dependence of the rates. The two aforementioned features of the stochastic processes, studied in this series, result in the corresponding symmetries of the twisted master operators $\hat{H}_{{\bm\chi}}$, more specifically, gauge invariance and Hermitian nature, respectively. These two symmetries of $\hat{H}_{{\bm\chi}}$, which have important implications for the properties of the generating function, were already outlined in section~\ref{generating-twisted} for the general case of an arbitrary network. In this subsection we discuss the gauge invariance and Hermitian of the twisted master operators in more detail and more explicitly for the case of our simple toy model.

\subsubsection{Current conservation and gauge invariance}
\label{toy-gauge-invariance}

The node populations cannot increase with time indefinitely, hence the currents, generated over a large number of cycles of the driving protocol, must satisfy the condition that the difference of such currents entering and leaving each of our two nodes is zero. If the currents through both links of our two-node model were independent, we should have to introduce a two-component vector ${\bm \chi} = (\chi_1,\chi_2)$ of the counting parameters and a more general twisted master operator
\begin{equation}\hat{H}_{\bm\chi}=\left(
\begin{array}{ll}
-\kappa_{1}(g_{1}+g_{2})& \kappa_{2}(g_{1}e^{-i\chi_1}+g_{2} e^{i\chi_2})\\
 \kappa_{1}(g_{1}e^{i\chi_1}+g_{2}e^{-i\chi_2}) & -\kappa_{2}(g_{1}+g_{2})
\end{array}\right),
\label{ham-chi22}
\end{equation}
compared to the one given by Eq.~(\ref{ham-chi2}). However, due to current conservation, the evolution with the operator given by Eq.~(\ref{ham-chi22}) contains the same information
 in the long-time limit as the simpler evolution operator of [Eq.~(\ref{ham-chi2}). This can be
justified by the fact that the eigenvalue $z_{{\bm\chi}}$ of the evolution operator $\hat{H}_{\bm\chi}$ is gauge invariant, i.e., it stays unchanged under the gauge transformations. A gauge transformation $\chi'_1=\chi_1 - \varphi_1 +\varphi_2$ and simultaneously $\chi'_2=\chi_2 - \varphi_2 +\varphi_1$, parametrized by the variables $\varphi_1$ and $\varphi_2$,  leaves $z_{{\bm \chi}}$ invariant, i.e.
\begin{equation}
\label{gauge-invariance2} z_{{\bm\chi}'}=z_{{\bm\chi}},
\end{equation}
which follows from the fact that
\begin{equation}
\hat{H}_{\bm \chi'}=e^{i\hat{\varphi}}
\hat{H}_{\bm \chi}
e^{-i\hat{\varphi}},
\label{ham-chi23}
\end{equation}
with
\begin{equation}
e^{i\hat{\varphi}}\equiv\left(
\begin{array}{ll}
e^{i\varphi_1}& 0\\
0 & e^{i\varphi_2}
\end{array}\right), \quad
e^{-i\hat{\varphi}} \equiv
\left(
\begin{array}{ll}
e^{-i\varphi_1}& 0\\
0 & e^{-i\varphi_2}
\end{array}\right).
\label{varphi-hat}
\end{equation}
The transformation, given by Eq.~(\ref{ham-chi23}), does not change the eigenvalues of the evolution operator
[Eq.~(\ref{define-U-twisted1})], since if $\vert u \rangle$ is an eigenvector of the evolution operator with the twisted master operator $\hat{H}_{{\bm\chi}'}$ [Eq.~(\ref{ham-chi23})], then $ e^{-i\hat{\varphi}} \vert u \rangle$ is an eigenvector of the evolution operator $\hat{H}_{{\bm\chi}'}$ [Eq.~(\ref{ham-chi22})] with the same eigenvalue. This gauge symmetry allows the parameters $\varphi_1$ and $\varphi_2$ to be chosen in a way that fixes the gauge with $\chi_2' =0$.

\subsubsection{Detailed balance and Hermitian property of twisted master operators}
\label{toy-hermitian}

When detailed balance is imposed, i.e., the rates satisfy the Arrhenius parametrization and are time-independent, no average currents are generated in a steady state. In fact, all odd cumulants of the current distribution must be also zero due to the time-reversal symmetry, which is guaranteed by the detailed balance conditions.

In the regime, in which all odd cumulants are zero, the generating function should be purely real, meaning that the eigenvalues of the twisted master operators with  kinetic rates constrained by  detailed balance must be also real. In fact, one can check directly that both eigenvalues of the operator, given by Eq.~(\ref{ham-chi2}), are real-valued functions of $\chi$.

At this point, we recall that the eigenvalues of Hermitian operators, which play a crucial role in
quantum mechanics, are all real. Apparently, the matrix in Eq.~(\ref{ham-chi2}) does not look Hermitian in the sense that its transpose does not coincide with its complex conjugate. Nevertheless, a simple transformation \cite{vankampen}
\begin{equation}
\hat{H}_{\bm \chi}^{h}=e^{\beta \hat{E}/2}
\hat{H}_{\bm \chi}
e^{-\beta \hat{E}/2}
\label{ham-herm}
\end{equation}
makes it Hermitian, and if $|u \rangle$ is the eigenvector of the Hermitian matrix $\hat{H}_{{\bm\chi}}^{h}$ [Eq.~(\ref{ham-herm})], then $e^{-\beta \hat{E}/2}| u \rangle $ is the eigenvector of the twisted
master operator $\hat{H}_{{\bm\chi}}$  [Eq.~(\ref{ham-chi2})], and the eigenvalues of both operators are identical. In other words, one can say that the twisted master operator is Hermitian with respect to a scalar product, weighted by set of additional energy-dependent factors [Eq.~(\ref{weighted-sclar-populations})].

\subsection{Topological invariants of full counting statistics}
\label{topo}

Depending on the values of the parameters, the twisted master operator in Eq.~(\ref{ham-chi2}) may have degenerate eigenvalues. It is a straightforward exercise to show that eigenvalue degeneracy happens when three
conditions are simultaneously satisfied \cite{sinitsyn-10jstat}, namely,
\begin{equation}
E_1=E_2, \quad W_1=W_2, \quad \chi=\pi.
\label{degen}
\end{equation}

Applying the language developed in the first manuscript of the series we say that parameters are {\it bad} if they satisfy conditions $E_{1}=E_{2}$ together with $W_{1}=W_{2}$. Otherwise we will say that parameters are
{\it good}. We recall that the space of good parameters is denoted ${\cal M}=\widetilde{{\cal M}}\setminus D$, where the discriminant space $D$ is the bad parameters. For our simple toy model the space ${\cal M}^{r}=\widetilde{{\cal M}}\setminus D^{r}$ of robust parameters, also introduced in the first manuscript of the series coincides with the space of good parameters. Since the good parameter space ${\cal M}$ and its discriminant subspace $D$ are $4$- and $2$-dimensional, respectively, a generic driving protocol avoids the bad parameter space, and generally one needs certain symmetry in the system for a driving protocol to go through the double degeneracy point, other than by just an accident.

In this subsection, we show that the full counting statistics of currents in our 2-node-2-link model, driven along a closed path in the good parameter space, can be classified using an integer number.

Since the twisted master operator is periodic in $\chi$, we can view $\chi$ as a parameter that resides in a cycle $[0,2\pi]$. Due to the periodic nature of driving, we can also think about the dimensionless time, $\tau$, as a variable that parameterizes another cycle $[0,1]$. Consider a torus, $T^{2}=S^{1}\times S^{1}$, parametrized by $(\chi,\tau)$ with the points $\chi=0$ and $\chi=2 \pi$, as well as $\tau=0$ and $\tau=1$ identified. With each point on the torus, we can associate an eigenvector, $|u_{\chi}(\tau)\rangle $, of the operator $\hat{H}_{\chi}$ that corresponds to the larger eigenvalue $\varepsilon_{\chi}(\tau)$, i.e.,
\begin{equation}
\hat{H}_{\chi}(\tau)|u_{\chi}(\tau)\rangle = \varepsilon_{\chi}(\tau)|u_{\chi}(\tau)\rangle.
\label{ev-chi2}
\end{equation}
The eigenstate choice is not unique because the multiplication of the eigenvector by a complex nonzero factor $\zeta_{\chi}(\tau)$ provides a valid eigenvector.

Consider a space, referred to as a {\it fiber bundle}, that is made of the {\it base} space represented by  the torus, $T^{2}=S^{1}\times S^{1}$, with each point $(\chi,\tau)\in T^{2}$ of the base equipped with a {\it fiber}, represented by complex nonzero factors $\zeta_{\chi}(\tau)$ that describe the freedom in the choice of an eigenvector. During adiabatic evolution of parameters, the twisted population vector that satisfies the twisted mater equation [i.e., the one obtained by replacing the master operator $\hat{H}(\tau)$ with its twisted counterpart $\hat{H}_{\chi}(\tau)$] follows the path $|u_{\chi}(\tau)\rangle$ of the instantaneous eigenstates at the instant values of $(\chi,\tau)$. Since we chose the contour to be always in the space of good parameters, this evolution leads to a unique continuous evolution of factors $\zeta_{\chi}(\tau)$, i.e., for a fixed $\chi$,  this evolution corresponds to some curve in the fiber bundle. The integer nature of the number of jumps requires the argument $\chi$ of the generating function to be a periodic variable, in particular $\zeta_{\chi+2\pi}=\zeta_{\chi}$. It is known that fiber bundles with continuous fiber fields $\zeta_{\chi}(\tau)$, taking values in nonzero complex numbers, are characterized by the {\it first Chern class}, and in the case of the base being a compact oriented $2$-dimensional manifold, e.g., a torus $T^{2}$ the first Chern class is represented by an integer number, hereafter referred to as the Chern number.

In our case the Chern number has a simple and transparent interpretation. After a large number of driving cycles the information on the initial state of our stochastic system is lost and the vector of generating functions satisfies the relation, $|u_{\chi}(\tau+1)\rangle =z_{\chi}|u_{\chi}(\tau) \rangle$, where we denote $z_{\chi}=\zeta_{\chi}(1)\ne 0$. Due to the periodicity in $\chi$, we can classify this function according to how many times the curve $z_{\chi}$ winds around the zero point in the complex plane while parameter $\chi$ grows from zero to $2\pi$. The Chern number in our model corresponds merely to this
winding number.


Having established that the counting statistics of currents in periodically driven systems can be classified by topological invariants,  the next questions are:

(i) How to calculate such invariants without solving a time-dependent
twisted master equation explicitly?

(ii)  What are the measurable manifestation of topological invariants in
stochastic processes?

We will address these two questions, for the toy model, in the following two subsections.

\subsection{Derivation of the winding number}
\label{toy-winding-derive}

In this subsection we establish a connection between the geometric phase of the generating function for our toy model and the winding number, and compute the latter using its topological nature by applying the low-temperature limit.

\subsubsection{ Geometric phase and winding number}
\label{toy-geom-ph-wind-numb}

In \cite{sinitsyn-07epl,sinitsyn-09review} it was shown that in the adiabatic limit the cumulant generating function (CGF) $\omega_{\chi} $ can be written as a sum of two contributions,
\begin{equation}
\label{z-om} \omega_{\chi}  = \omega_{\chi}^{g}+\omega_{\chi}^{q}, \;\;\; z_{\chi}=z_{\chi}^{g}z_{\chi}^{q}
\end{equation}
referred to as the geometric and quasisteady terms. The quasisteady contribution has a natural form
\begin{equation}
\label{qst-tr} \omega_{\chi}^{q} = \int_{0}^{\tau_{D}}dt\varepsilon_{\chi}(t)= \tau_{D}\int_{0}^{1}d\tau \varepsilon_{\chi}(\tau).
\end{equation}
The geometric contribution $\omega_{\chi}^{g}$ depends only on the geometry of the contour  in the parameter space ${\cal M}$ (which justifies the notation), and is independent of the rate of motion along this contour, as long as the adiabatic approximation can be employed. It has the form
\begin{equation}
\omega_{\chi}^{g} = -\int_{\bm s}{\bm A}(\chi,{\bm x})\cdot d{\bm x}, \quad
{\bm A}(\chi,{\bm x}) =\langle u_{\chi}({\bm x})|\partial_{\bm x}u_{\chi}(\bm x)\rangle,
\label{sgeom-tr}
\end{equation}
or equivalently
\begin{equation}
\omega_{\chi}^{g} = -\int_{0}^{1}d\tau A_{0}(\chi,\tau), \quad A_{0}(\chi,\tau)=\langle u_{\chi}(\tau)|\partial_{\tau}u_{\chi}(\tau)\rangle \, .
\label{sgeom-tr-pull-back}
\end{equation}
Introducing also
\begin{equation}
\label{A-other-component} A_{1}(\chi,\tau)=\langle u_{\chi}(\tau)|\partial_{\chi}u_{\chi}(\tau)\rangle,
\end{equation}
we can interpret ${\bm A}=(A_{0},A_{1})$ as a vector potential (or equivalently an abelian gauge field) defined on our torus. The reason why ${\bm A}$ is referred to as a gauge field originates from the fact that the ground state $|u(\chi,\tau)\rangle$ is defined up to a scalar factor $\zeta_{\chi}(\tau)$, and a rescaling $|u'_{\chi}(\tau)\rangle= \zeta_{\chi}(\tau)|u_{\chi}(\tau)\rangle= e^{\phi(\chi,\tau)}|u_{\chi}(\tau)\rangle$ leads to a gauge transformation of the vector potential
\begin{equation}
\label{toy-gauge-transf} A'_{j}=A_{j}-\partial_{j}\phi, \quad |u'\rangle=e^{\phi}|u\rangle
\end{equation}
where $\partial_{0}=\partial_{\tau}$ and $\partial_{1}=\partial_{\chi}$,

In terms of the cumulant generating function (CGF), the topological invariant has a new physical interpretation: while the generating function $ z_{\chi}$ is a periodic function of $\chi$, the CGF, $\omega_{\chi}$, may not be necessarily periodic, but it rather has the property, $\omega_{\chi+2\pi} =\omega_{\chi} +2\pi i Q$, where $Q$ is the integer that coincides with the winding number. The CGF is a measurable function. By measuring sufficiently many cumulants of the distribution, one can determine $Q$ if the corresponding representation of the CGF in terms of a series of cumulants converges for arbitrary $\chi\in [0,2\pi]$.

The geometric contribution does not appear in steady states, since it is nonzero only when the parameters are varying along the contour ${\bm s}$. The steady state CGF would be given by $\varepsilon_{\chi}\tau_D$, which means that $\omega_{\chi}^{q} $  is just the time average steady state contribution. The eigenvalues of the twisted master equation are real because of the Hermitian property of the twisted master operator at detailed balance conditions on the kinetic rates. This means that  the strict quasisteady state contribution to the CGF is purely real. In particular, it contributes only to the even cumulants of the currents and is not related to the winding number.  Thus the information about the topological invariant is contained in the geometric contribution,  $\omega_{\chi}^{g} $, specifically, the winding number, $Q$, given by
\begin{equation}
\label{winding} Q=\frac{1}{2\pi i}\int_{0}^{2\pi}d\chi\partial_{\chi}\omega_{\chi}^{g}=\frac{1}{2\pi i}\int_{0}^{2\pi}d\chi\frac{\partial_{\chi}z_{\chi}^{g}}{z_{\chi}^{g}}.
\end{equation}
Note that since $\omega_{\chi}^{q}$ is a single-valued function of $\chi$, one can replace $\omega_{\chi}^{g}$ and $z_{\chi}^{g}$ in Eq.~(\ref{winding}) with $\omega_{\chi}$ and $z_{\chi}$, respectively.

There is a delicate and important detail that needs to be discussed. The ground state $|u_{\chi}(\tau)\rangle$ is globally defined on the torus $T^{2}$ only up to a scalar position-dependent factor $\zeta_{\chi}(\tau)$ that may or may not be fixed globally in a smooth (or even continuous) way. Actually, the nonzero value of the winding number $Q$ forms a topological obstruction to fixing $\zeta_{\chi}(\tau)$ globally. Therefore, as explained in some detail in appendix~\ref{Chern-class}, the vector potential ${\bm A}=(A_{0},A_{1})$ is defined globally only up to a gauge transformation [Eq.~(\ref{toy-gauge-transf})]. More precisely this means that the vector potential can be defined in some regions that cover our torus, so that in the intersection of any two regions the vector potentials differ by a gauge transformation. Therefore, the expression for $\omega_{\chi}^{g}$ [Eq.~(\ref{sgeom-tr-pull-back})] should be understood in the following way. Fixing $\chi$ we can define $|u_{\chi}(\tau)$ together with the scalar factors on a cycle $0\le \tau \le 1$ and further apply Eq.~(\ref{sgeom-tr-pull-back}). Being represented by an integral of the vector potential over a closed contour, $\omega_{\chi}^{g}$ is gauge invariant and therefore well-defined.


It is still possible to come up with an integral representation of the winding number $Q$ with a gauge-invariant integrand. To that end we introduce the curvature (``magnetic field'') $F$ of the vector potential
\begin{equation}
\label{berr-curr} F(\chi,\tau)= \partial_{0}A_{1}-\partial_{1}A_{0}= \langle \partial_{\tau}u| \partial_{\chi}u\rangle - \langle \partial_{\chi}u|\partial_{\tau} u
\rangle,
\end{equation}
sometimes referred to as the Berry curvature \cite{sinitsyn-09review}, so that, as justified with some detail in appendix~\ref{Chern-class}, Eq.~(\ref{winding}) can be recast as
\begin{equation}
\label{winding25} Q= \frac{1}{2\pi}\int_0^{2\pi}d\chi\int_0^1 d\tau F(\chi,\tau),
\end{equation}
with the curvature, due to its gauge-invariant nature, being a smooth function globally defined on our torus $T^{2}$. Also note that the r.h.s. of Eq.~(\ref{winding25}) actually reproduces the original definition of the first Chern class for $2$-dimensional manifolds, so that Eq.~(\ref{winding25}) provides a Chern number view of the topological invariant $Q$ of the full counting statistics.

A calculation of our topological invariant via Eq.~(\ref{winding25}) can still be technically a hard problem. It requires knowing the eigenstates of the twisted master operators, that might not typically be found analytically. Fortunately, for stochastic models there is an alternative procedure.

A key observation that makes calculation of the winding number in our toy model easy is that the degeneracy conditions [Eq.~(\ref{degen})] do not depend on the temperature, so if we consider a continuous class of periodically driven systems that differ only by temperature, the winding numbers that characterize each of these models should be the same because continuous changes of temperature can lead only to continuous changes of $\zeta_{\chi}(\tau)$, that cannot change topological invariants of the fiber bundle. Thus we can, in fact, calculate the winding number in the limit of zero temperature, $\beta\to \infty$.

\subsubsection{Winding number in the low-temperature limit}
\label{toy-wind-low-temp}

Taking the low-temperature limit leads to considerable simplifications since the ground state of the master operator can be found explicitly at each point $(\chi,\tau)\in T^{2}$ of the torus up to exponentially suppressed
corrections. For our two-state model, whenever the barriers, $W_1$ and $W_2$ are different, we can disregard the factors which contain $e^{-\beta W_1}$ or $e^{-\beta W_2}$, if $W_1>W_2$ or $W_2>W_1$, respectively. Still, there can be short time intervals when the driving protocol passes through the point $ W_1 = W_2$. Precisely at the latter intervals, the restriction of dealing only with those protocols that lie entirely in the good parameter space leads to the condition $E_1 \ne E_2$, and hence we can disregard terms containing $e^{\beta E_1}$ or $e^{\beta E_2}$, if, respectively, $E_1<E_2$ or $E_2<E_1$.

Consider again the driving protocol, $E_2=W_2=0$, $E_1(\tau)=\cos(2\pi \tau)$,  $W_1(\tau)=\sin(2\pi \tau)$, that has been already explored in the first manuscript of the series to study average currents. One can distinguish four time segments with distinct behavior of the system:
\medskip

\noindent (a) for $\tau\in [-\varepsilon,\varepsilon]$, where $\varepsilon< \pi/2$, the path goes through the point $W_1=W_2$ but we always have $E_1>E_2$;
\smallskip

\noindent (b) for $\tau \in (\varepsilon,\pi-\varepsilon)$, we have $W_1>W_2$;
\smallskip

\noindent (c) for $\tau\in [\pi-\varepsilon,\pi+\varepsilon]$, we pass through the point $W_1=W_2$ but we always have $E_1<E_2$;
\smallskip

\noindent(d) for $\tau\in [\pi+\varepsilon,2\pi-\varepsilon]$, we have $W_2>W_1$.
\medskip

Consider stochastic dynamics on each of those segments in the low-temperature $\beta\to\infty$ limit. At segment (a), we can disregard the terms with the factor $e^{\beta E_2}$, so that
\begin{equation}
\label{ham-chi2-i}
\hat{H}_\chi^{(a)} \approx \left(
\begin{array}{ll}
-\kappa_{1}(g_{1}+g_{2})& 0\\
\kappa_{1}(g_{1}e^{i\chi}+g_{2}) & 0
\end{array}\right),
\end{equation}
has a right eigenvector
\begin{equation}
|u^{(a)} \rangle= \left(
\begin{array}{l}
0\\
1
\end{array}\right)
\label{ev-1}
\end{equation}
that has a simple interpretation: the fact that it does not depend on time means that it represents a process during which the system remains in the second node without producing any current.

By analogy with (a), the evolution at segment (c) is also trivial, and it is represented by the eigenstate
\begin{equation}
|u^{\rm (c)}\rangle= \left(
\begin{array}{l}
1\\
0
\end{array}\right),
\label{ev-3}
\end{equation}
indicating that the system stays at the first node for the entire time segment.

At segment (b), we can disregard the terms with the factor $e^{-\beta W_1}$. The approximate twisted master
operator, for  $W_2<W_1$, turns out to be independent also of the counting parameter $\chi$:
\begin{equation}
\hat{H}_{\chi}^{(b)}=\left(
\begin{array}{ll}
- e^{\beta( E_1-W_2)}& e^{\beta (E_2-W_2)}\\
e^{\beta (E_1-W_2)} & -e^{\beta (E_2-W_2)}
\end{array}\right).
\label{ham-tr2}
\end{equation}
It is equivalent to the non-twisted master operator of the model in
Fig.~\ref{2level-split}(b). The largest eigenvalue of such an operator is zero.
The right and left eigenstates of $H_{\chi}^{(b)}$ that correspond to the zero eigenvalue are given by
\begin{equation}
|u^{(b)}\rangle=Z^{-1}\left(
\begin{array}{l}
e^{-\beta E_1}\\
e^{-\beta E_2}
\end{array}\right), \quad \langle u^{(b)}|=(1,1),
\label{ev-tr2}
\end{equation}
with $Z=e^{-\beta E_1}+e^{-\beta E_2}$ being the the normalizing factor. Generally, the eigenstates of the operator given by Eq.~(\ref{ham-tr2}) can be chosen with an arbitrary phase $e^{\varphi (\chi,{\bm x})}$. However, in order to apply Eq. (\ref{sgeom-tr-pull-back}), we should choose the eigenstates of the twisted master operator so that they continuously change along the contour ${\bm s}$. The choice of the real prefactor in Eq.~(\ref{ev-tr2}) guarantees that at the points where the segments (a) and (c) overlap with (b), the eigenstates coincide up to exponentially suppressed terms which may be disregarded.

\begin{figure}
\centerline{\includegraphics[width=2.8in]{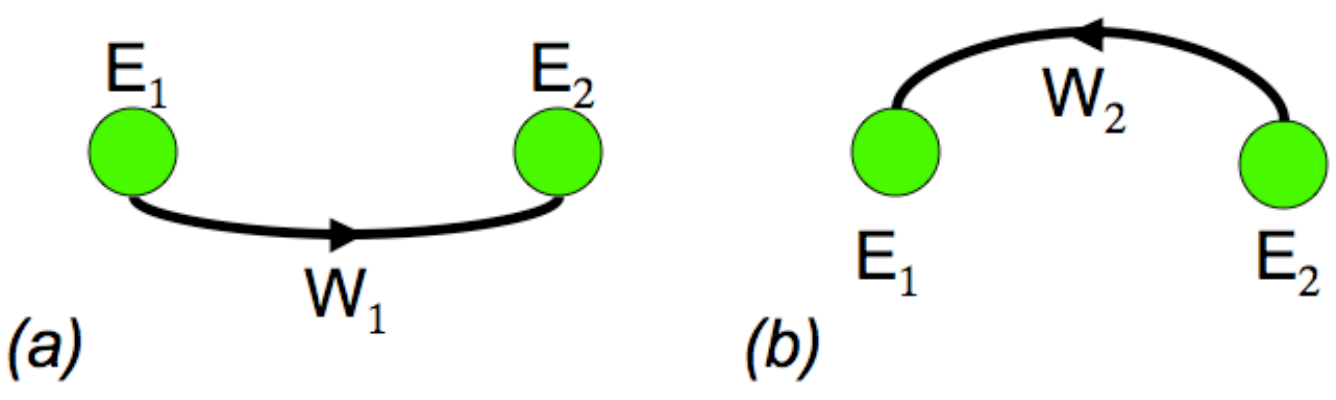}}
  \caption{ Two steps (a) $W_1<W_1$ and (b) $W_1>W_2$. At  the low temperature limit, the effective model is a diffusion on a tree graph that consist only of one link and two sites.}
\label{2level-split-1}
\end{figure}

Finally, at segment (d) the terms that contain $e^{-\beta W_2}$ can be disregarded, so that
\begin{equation}
\hat{H}_{\chi}^{(d)}=\left(
\begin{array}{ll}
- e^{\beta (E_1-W_1)}& e^{\beta (E_2-W_1)-i\chi}\\
e^{\beta (E_1-W_1)+i\chi} & -e^{\beta (E_2-W_1)}
\end{array}\right).
\label{ham-tr1}
\end{equation}
The matrix in Eq.~(\ref{ham-tr1}) corresponds to the twisted master operator for the Markov chain model illustrated in Fig.~\ref{2level-split-1}(a) that has  two sites and one link only. Such a matrix should have the lowest eigenvalue independent of $\chi$, because otherwise the tree graph would be able to support a steady state current with growing cumulants, which is impossible. Since $\chi=0$ corresponds to no twisting, the aforementioned largest eigenvalue should be exactly zero. The above intuitive arguments can be put on firm ground by noticing that the twisted operator in Eq.~(\ref{ham-tr1}) can be transformed to an untwisted counterpart by a gauge transformation in a form of Eq.~(\ref{ham-chi23}). The eigenstates of $H_{\chi}^{(d)}$ corresponding to the zero eigenvalue are
\begin{equation}
\vert u^{(d)}  \rangle=Z^{-1}e^{i\phi} \left(
\begin{array}{l}
e^{-\beta E_1}\\
e^{-\beta E_2+i\chi}
\end{array}\right), \quad \langle u^{(d)}  \vert = e^{-i\phi} (1, e^{-i\chi} ).
\label{ev-4}
\end{equation}
The phase $\phi =\phi(\chi,\tau)$ should be chosen in a way such that $|u_{\chi}(\tau)\rangle$ is continuous at $\tau=2\pi-\varepsilon$ and $\tau=\pi+\varepsilon$ that corresponds to the intersection points of segment (d) with the segments (a) and (c), respectively. This requires  $\phi(\chi,2\pi-\varepsilon)=0$ and $\phi(\chi,\pi+\varepsilon)=-\chi$, respectively.

In such a gauge, the eigenvectors smoothly change from one interval into another. Calculating the relevant component $A_{0}$ of the vector potential,  we find that in our gauge $A_{0}(\tau)=0$ at the segments  (a), (b), and (c). At segment (d), we find
\begin{equation}
A_{0}=i\partial_{\tau}\phi.
\label{gauge}
\end{equation}
Performing integration along segment (d), we find
\begin{eqnarray}
\omega_{\chi}^{g}&=&-\int_{\pi+\varepsilon}^{2\pi-\varepsilon}d\tau A_{0}(\tau) \nonumber \\ &=&-i\left(\phi(\chi,2\pi-\varepsilon)-\phi(\chi,\pi+\varepsilon)\right)=i\chi.
\label{sgeom-tr2}
\end{eqnarray}
We conclude that in the low temperature limit the geometric phase contributes to the first cumulant of the current statistics only, provided that the exponentially suppressed corrections are neglected. Another observation is that $\omega_{\chi}^{\rm g} $ is not a periodic function of $\chi$, and the winding number is equal to one:
\begin{equation}
Q=1.
\label{winding3}
\end{equation}

The above arguments can be extended to a general periodic protocol in a straightforward way. For our toy model, we can always split any periodic driving protocol into alternating time segments such that (a) $W_1\approx W_2$ while $E_1<E_2$, (b) $W_1>W_2$, (c) $W_1\approx W_2$ while $E_2<E_1$, and (d) $W_1<W_2$. The continuous choice of the eigenstates in each of those segments is the same as in the above example. Then whenever we find the path in the parameter space that goes from (c) to (a) via (d), the geometric contribution to CGF acquires a term $i\chi$ and whenever the path goes from (a) to (c) via (d) the CGF acquires a term $-i\chi$. So the CGF per period of a driving is given by
\begin{eqnarray}
\label{genf-end}
\omega_{\chi}=iQ\chi+\int_{0}^{\tau_D}dt\varepsilon_{\chi}(t)
=iQ\chi+\tau_{D}\int_{0}^{1}d\tau\varepsilon_{\chi}(\tau),
\end{eqnarray}
where $Q$ is the integer number, which is equal to the number of transitions through the interval (d) taken with a proper sign as discussed above.

We already mentioned that $\varepsilon(\chi,t)$ is purely real and contributes to the even cumulants only. The coefficient of $\chi^{n}$ in the Taylor expansion of the CGF corresponds to $n$-th cumulant of the corresponding probability distribution. Our results show that in the $\beta\to\infty$ limit the geometric contribution is linear in $\chi$. Therefore, in addition to identifying the first cumulant, i.e., average current with the Chern number $Q$, Eq.~(\ref{genf-end}) also demonstrates vanishing of all higher odd cumulants of the generated current in the low-temperature limit.

In the following sections, we will show that the quantized current in a generic network is also a manifestation of topological properties of the full counting statistics of currents in periodically driven networks.

The main difference between the toy model and a generic network is that the latter typically has more than one independent current, i.e., the argument of the generating function is typically multi-variant. Also, for a generic graph, the condition for the largest eigenvalue degeneracy is typically temperature dependent. However, at sufficiently low, yet finite temperature, the degeneracy can be avoided for all values of the CGF argument ${\bm\chi}$, provided the parameters belong to the good parameter space. Indeed, by restricting the twisted master operator to the minimal spanning tree, which can be justified by the same arguments as for steps (b) and (d) in the  toy model, we find that the largest eigenvalue is zero and non-degenerate. Therefore, if the temperature is finite, but small enough, the largest eigenvalue is still non-degenerate, provided the parameters are good, and the topological invariants, obtained by considering deterministic evolution in the context of average currents, characterize the full counting statistics, at least in some range of temperatures.

\subsection{Toy model for fractional quantization}
\begin{figure}
\centerline{\includegraphics[width=2.8in]{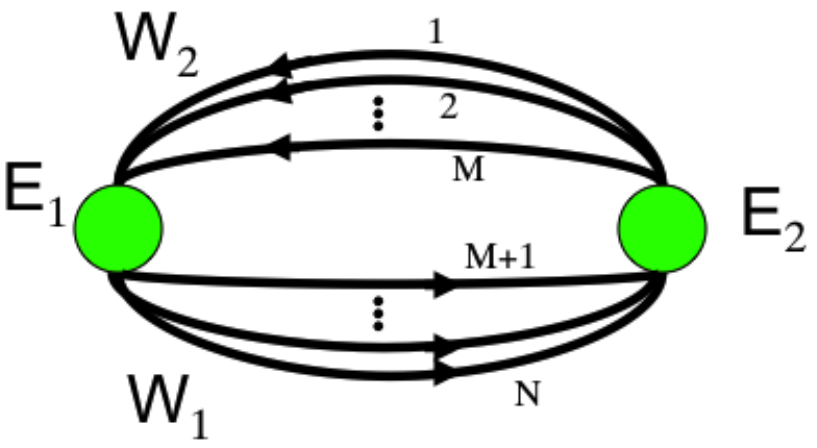}}
  \caption{ A graph with 2 nodes and $N$ links. Nodes are
    characterized by energies $E_1$ and $E_2$. There are two sets of
    links that are permanently degenerate, so that links with numbers
    $1,\ldots M$  have the same barrier size $W_2$, and the links
    with numbers $M+1, \ldots N$ have barrier size $W_1$. }
\label{N-level_pic}
\end{figure}

According to \cite{sinitsyn-09jcp}, when permanent degeneracies of the node energies and/or barrier heights take place, the current, generated by adiabatic periodic evolution of the system parameters, is generally fractional, rather than integer-valued in the low-temperature limit. In this subsection, we consider a simple model that only slightly generalizes the model, considered so far in this section (see Fig.~\ref{2level-1}), by replacing each of two links by a group of links with the permanently degenerate barrier heights, as illustrated in Fig.~\ref{N-level_pic}. Let $g_{\alpha}=e^{-\beta W_{2}}$ for $\alpha=1,\ldots M$, and $g_{\alpha}=e^{-\beta W_{1}}$ for $i=M+1,\ldots N$. We also introduce a vector ${\bm \chi} = (\chi_1, \ldots \chi_N)$ of counting parameters that monitor the number of counterclockwise minus clockwise transitions through each link. The evolution operator for our $N$-component moment generating function is given by
\begin{widetext}
\begin{equation}
\hat{H}_{\bm \chi}=\left(
\begin{array}{cc}
-\kappa_{1}\sum \limits_{i=1}^N g_{i}& \kappa_{2}(\sum \limits_{i=1}^Mg_{i}e^{i\chi_i}+\sum \limits_{j=M+1}^Ng_{j}e^{-i\chi_j})\\
\kappa_{1}(\sum \limits_{i=1}^Mg_{i}e^{-i\chi_i}+\sum \limits_{j=M+1}^Ng_{j}e^{i\chi_j}) & -\kappa_{2}\sum \limits_{i=1}^N g_{i}
\end{array}\right).
\label{ham-chi33}
\end{equation}
\end{widetext}

\begin{figure}
\centerline{\includegraphics[width=2.8in]{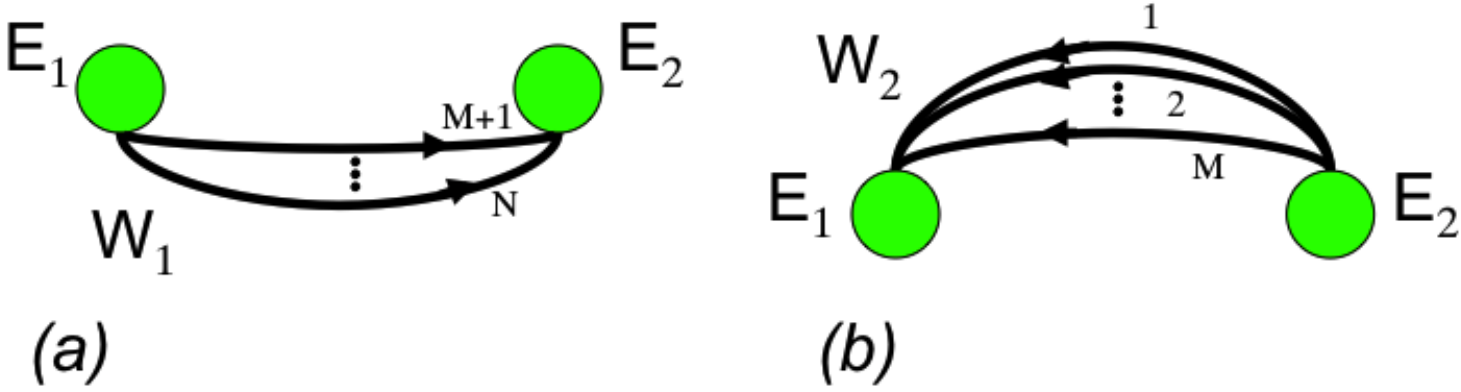}}
  \caption{Effective kinetic models that describe evolution of the system
    in Fig.~\ref{N-level_pic} in the limit of low temperature for (a) the
    case of $W_{1}<W_2$ and (b) the case of $W_1>W_2$. }
\label{network-steps2}
\end{figure}

In the adiabatic limit, the calculation of the CGF is reduced to the eigenvalues and eigenvectors of the 2$\times$2 matrix $\hat{H}_{\bm \chi}$ in Eq.~(\ref{ham-chi33}), which can be completed explicitly. Moreover, in the low-temperature limit, one can split the driving protocol period time into time segments intervals (a)-(d), similar to what has been done for the  2-state-2-link model, and apply similar approximations, based on disregarding the links with the higher barriers, when computing the generating function, as illustrated in Fig.~\ref{network-steps2}. For the same driving protocol, as considered for the 2-state-2-link model earlier in this section, the calculation, described above, results in
\begin{equation}
z_{\bm \chi}(\tau_{D}) = \left( \frac{\sum_{j=1}^Me^{i\chi_j} }{M}
\right)
\left( \frac{\sum_{j=M+1}^Ne^{i\chi_j} }{N-M} \right)
e^{\epsilon({\bm \chi},\tau_{D})},
\label{genNM}
\end{equation}
where $\epsilon({\bm \chi},\tau_{D})$ is a non-universal quasisteady-state contribution that is purely real with $\epsilon({\bm\chi},\tau_{D})=\epsilon(-{\bm \chi},\tau_{D})$, hence obviously contributes to the even cumulants only, and, in this sense, is indistinguishable from the thermal noise at equilibrium. The most interesting feature of the counting statistics arises from the pre-exponential factor that describes the effect of the geometric phase. When the driving protocol encloses the degeneracy point in the parameter space $Q$ times, the geometric factor in $z_{\bm \chi}$ becomes
\begin{equation}
z_{\bm \chi}^{g}(\tau_D)=\left( \frac{\sum_{j=1}^{M}e^{i\chi_j} }{M}\right)^Q\left(\frac{\sum_{j=M+1}^{N}e^{i\chi_j}}{N-M} \right)^Q.
\label{genNM2}
\end{equation}
We reiterate that the contributions to the odd cumulants are provided by the geometric factor [Eq.~(\ref{genNM2})] only. The components of the average generated currents can be obtained by taking the derivatives of $\ln\left(z_{\bm \chi}^{g}(\tau_D)\right)$  with respect to the corresponding counting parameters $i\chi_{j}$, using Eq.~(\ref{genNM2}), followed by setting ${\bm\chi}=0$ to zero:
 \begin{equation}
\langle Q_{i} \rangle = Q/M,\quad {\rm if} \quad {i=1,\ldots M},
\label{QNM1}
\end{equation}
and
\begin{equation}
\langle Q_{i} \rangle = Q/(N-M),\quad {\rm if} \quad {i=M+1,\ldots N}.
\label{QNM2}
\end{equation}
As expected, at low temperatures the average currents through individual links are  fractionally
quantized. The fractional quantization is expressed through the integer parameters, $Q$, $N$ and $M$. The latter two originate from degeneracy numbers  for the two groups of links. The integer number $Q$ is the winding number, which still has a Chern class interpretation. This can be seen from the fact that if we set
$\chi_1=\ldots=\chi_M=0$ and $\chi_{M+1}=\ldots=\chi_N=\chi$ the twisted master operator in Eq.~(\ref{ham-chi33}) is the same as in the 2-state-2-link model, in which $Q$ has a Chern number interpretation.

We conclude this section with noting that we explicitly calculated the CGF $\omega_{{\bm\chi}}$ or related quantity $z_{{\bm\chi}}=e^{\omega_{{\bm\chi}}}$ in the low-temperature adiabatic limit for two toy models, both containing two nodes. The result for the $2$-link model that supports integer quantization [Eq.~(\ref{genf-end})] looks universal, with the geometric contribution totally described by the topological invariant $Q$ represented by the first Chern class. Eq.~(\ref{genf-end}) suggests that the above universality should hold for a general network. Indeed, we will demonstrate in section~\ref{generic-networks} that this is the case. Therefore, the Chern class interpretation of the average integer-quantized current is a consequence of a fundamental property of the CFG in the case of no permanent degeneracy. The expression for the geometric factor [Eq.~(\ref{genNM2})] in the CFG of the extended toy model that supports fractional quantization is more complicated. Therefore, although in section~\ref{chern-II} we provide a Chern class interpretation of fractionally-quantized current, based on the averaging concept, formulated in the first manuscript of the series, the presented interpretation, as opposed to the integer case, is somewhat artificial and does not reflect any universal topological properties of the CGF, which are yet to be determined.

\section{General Networks}
\label{generic-networks}

In this section, we study the generating function on arbitrary graph with the focus on the
low-temperature adiabatic limit.

\subsection{Generating function in the adiabatic limit and topological contributions}
\label{generating-Chern-adiabatic}

We start with considering the adiabatic $\tau_{D}\to\infty$ limit followed by applying the low-temperature $\beta\to\infty$ limit. In what follows, we will be using the multiplicative representation ${\bm\lambda}$ for the multi-variant argument of the generating function. We reiterate that the components in the multiplicative and additive representations are related by $\lambda_{\alpha}= e^{i\chi_{\alpha}}$.

Adiabatic calculation of the eigenvalue $z_{{\bm\lambda}}({\bm s};\tau_{D},\beta)$ of the time-ordered exponential operator $\hat{U}_{{\bm\lambda}}({\bm s};\tau_{D},\beta)$ defined by Eq.~(\ref{define-U-twisted}) becomes standard when the ground state of the twisted master operator $\hat{H}_{\bm\lambda}({\bm s};\beta)$ is non-degenerate at all times. We reiterate that, as justified in  section~\ref{generating-twisted}, the notion of a ground state is well-defined due to Hermitian property of the twisted master operators $\hat{H}_{{\bm\lambda}}$, which restricts its eigenvalues to be real. To reproduce the complete generating function in this standard manner, we want this non-degeneracy to take place for values of the argument ${\bm\lambda}$. Stated differently, we would like our driving protocol to stay inside the parameter subspace ${\cal M}_{X}^{r}(\beta)\subset \widetilde{{\cal M}}_{X}$ (that naturally depends on $\beta$) characterized by the property that $\hat{H}_{\bm\lambda}({\bm x};\beta)$ has a non-degenerate ground state for all ${\bm\lambda}$, as long as ${\bm x}\in {\cal M}_{X}^{r}(\beta)$. The subspace ${\cal M}_{X}^{r}(\beta)$ will be referred to as the {\it non-degenerate ground-state space}, associated with $\beta$. The parameter sets ${\bm x}\in {\cal M}_{X}^{r}(\beta)$ that belong to this space and the driving protocols that lie entirely inside this space will be referred to as non-degenerate ground-state parameters and driving protocols, respectively.  Although at this point it would seem that the subspace ${\cal M}_{X}^{r}(\beta)$ has been introduced based on our ability to perform adiabatic calculation, we will show later that in the low-temperature limit the above subspace actually describes the robust driving protocols.

As justified above, for a non-degenerate ground-state driving protocol, by applying a standard adiabatic technique, we have in the $\tau_{D}\to\infty$ limit
\begin{eqnarray}
\label{z-adiabatic} z_{{\bm\lambda}}({\bm s};\tau_{D},\beta)&\approx& \zeta_{{\bm\lambda}}({\bm s};\beta)e^{\tau_{D}\Lambda_{\bm\lambda}^{A}({\bm s};\beta)}, \nonumber \\ \Lambda_{\bm\lambda}^{A}({\bm s};\beta)&\equiv& \int_{0}^{1}d\tau \omega_{{\bm\lambda}}({\bm s}(\tau);\beta),
\end{eqnarray}
and we reiterate that $\omega_{{\bm\lambda}}({\bm x};\beta)$ denotes the lowest eigenvalue of the twisted master operator $\hat{H}_{\bm\lambda}({\bm x};\beta)$, and $\zeta_{{\bm\lambda}}({\bm s};\beta)$ is the geometric factor. An attempt to represent the geometric contribution in  exponential form (similar to the second factor) faces a topological {\it obstruction}, represented by a set ${\bm n}({\bm s};\beta)=(n_{\alpha}({\bm s};\beta)|\alpha\in X_{1})$ of integers, where $n_{\alpha}({\bm s};\beta)$ is the index that measures how many times $\zeta_{{\bm\lambda}}({\bm s};\beta)$ winds around the zero point $0\in\mathbb{C}$, while $\lambda_{\alpha}$ goes over a unit circle $|\lambda_{\alpha}|=1$ (as explained in some detail in appendix~\ref{Chern-class}), whereas all other components of ${\bm\lambda}$ stay unchanged. However, we can still represent the geometric contribution
\begin{eqnarray}
\label{geometric-decompose} \zeta_{{\bm\lambda}}({\bm s};\beta)=\zeta_{{\bm\lambda}}^{T}({\bm s};\beta)e^{\Lambda_{{\bm\lambda}}^{G}({\bm s};\beta)}
\end{eqnarray}
as a product of a purely topological factor
\begin{eqnarray}
\label{topological-factor} \zeta_{{\bm\lambda}}^{T}({\bm s};\beta)=\prod_{\alpha\in X_{1}}\lambda_{\alpha}^{n_{\alpha}({\bm s};\beta)}
\end{eqnarray}
and an exponential term, where $\Lambda_{{\bm\lambda}}^{G}({\bm s};\beta)$ describes a purely geometric correction to the topological term. Summarizing, we have
\begin{eqnarray}
\label{z-adiabatic-2} z_{{\bm\lambda}}({\bm s};\tau_{D},\beta)\approx (\prod_{\alpha\in X_{1}}\lambda_{\alpha}^{n_{\alpha}({\bm s};\beta)})e^{\Lambda_{{\bm\lambda}}^{G}({\bm s};\beta)+\tau_{D}\Lambda_{{\bm\lambda}}^{A}({\bm s};\beta)},
\end{eqnarray}
and, due to gauge invariance
\begin{eqnarray}
\label{n-conservation} \sum_{j\ne i}\sum_{\alpha}^{\partial\alpha=\{i,j\}}n_{i\alpha}({\bm s};\beta)=0,
\end{eqnarray}
which means that the set ${\bm n}({\bm s};\beta)$ of integers represents a conserving integer-valued current.

The obtained adiabatic expression [Eq.~(\ref{z-adiabatic-2})] has the following important implication on the LD function. Let $\tilde{{\cal S}}({\bm Q})$ be the LD function that corresponds via the Legendre transform to
\begin{eqnarray}
\label{Lambda-ref} \tilde{\Lambda}_{\bm\lambda}({\bm s};\tau_{D},\beta)\equiv \Lambda_{\bm\lambda}^{G}({\bm s};\beta)+\tau_{D}\Lambda_{{\bm\lambda}}^{A}({\bm s};\beta).
\end{eqnarray}
Then
\begin{eqnarray}
\label{S-shift} {\cal S}({\bm Q})=\tilde{\cal S}\left({\bm Q}-{\bm n}({\bm s};\beta)\right),
\end{eqnarray}
i.e., the topological factor results in just a shift of the LD function and, therefore, affects the average current only.

We are now in a position to show that the integers $n_{\alpha}({\bm s};\beta)$ that describe the shift in Eq.~(\ref{S-shift}) can be viewed as Chern numbers, and, therefore, calculated using the
integral representation, given by Eq.~(\ref{Chern-number-2}). Indeed, by the definition of the robust parameter space ${\cal M}_{X}^{r}$, the family $\hat{H}_{\bm\lambda}({\bm x};\beta)$ of twisted master operators provides a linear fiber bundle ${\cal L}^{\beta}$ over ${\cal M}_{X}^{r}\times {\cal T}$. We reiterate that a linear fiber bundle is just a collection of fibers ${\cal L}_{{\bm x},{\bm\lambda}}^{\beta}$, parametrized by the points $({\bm x},{\bm\lambda})\in {\cal M}_{X}^{r}\times{\cal T}$, with the fiber ${\cal L}_{{\bm x},{\bm\lambda}}^{\beta}$ being a complex $1$-dimensional vector space represented by the ground states $|u\rangle$ that satisfy $\hat{H}_{\bm\lambda}({\bm x};\beta)|u\rangle=\omega_{\bm\lambda}({\bm x};\beta)|u\rangle$, with $\omega_{\bm\lambda}({\bm x};\beta)$ being the lowest eigenvalue.

We further observe that a periodic driving protocol ${\bm s}$ defines a set ${\bm f}({\bm s})=(f_{\alpha}({\bm s})|\alpha\in X_{1})$ of $2$-dimensional cycles in ${\cal M}_{X}^{r}\times {\cal T}$, associated with the network links. A cycle that maps a torus $T^{2}$ to ${\cal M}_{X}^{r}\times {\cal T}$ associates with a point $(\tau,\chi)$ in the torus (we reiterate that $0\le \tau\le 1$ and $0\le\chi\le 2\pi$ with periodic boundary conditions) the point $({\bm s}(\tau),{\bm\lambda}^{\alpha}(\chi))$, where the components of ${\bm\lambda}^{\alpha}(\chi)$ are: $\lambda_{\alpha}^{\alpha}(\chi)=e^{i\chi}$, and $\lambda_{\gamma}^{\alpha}(\chi)=1$ for $\gamma\ne \alpha$. Stated less formally, if we view a torus $T^{2}$ as a product of two circles, then the two circles of the cycle $f_{\alpha}({\bm s})$ are represented by the driving protocol and the unit circle $|\lambda_{\alpha}|=1$, respectively. Restricting the argument ${\bm\lambda}$ of the generating function to the circle $\lambda_{\gamma}=e^{i\chi\delta_{\gamma}^{\alpha}}$ we identify $n_{\alpha}({\bm s};\beta)=n[\zeta(\chi))]$ with the latter given by Eq.~(\ref{integral-border}). Finally, due to Eq.~(\ref{Chern-number-3})
\begin{eqnarray}
\label{index-to-Chern} n_{\alpha}({\bm s};\beta)=c_{1}({\cal L}^{\beta})f_{\alpha}({\bm s}),
\end{eqnarray}
which can be interpreted as that $n_{\alpha}({\bm s};\beta)$ is a Chern number, obtained by evaluating the first Chern class $c_{1}({\cal L}^{\beta})$ of ${\cal L}^{\beta}$ at a $2$-dimensional cycle $f_{\alpha}({\bm s})$. In particular it can be calculated by using the integral representation of Eq.~(\ref{Chern-number-2}).

\subsection{Low-temperature limit for generating function}
\label{generating-Chern-low-temp}

With our focus on the low-temperature limit, it is desirable to replace the ground-state non-degenerate temperature-dependent subspaces ${\cal M}_{X}^{r}(\beta)$ with some limiting subspace ${\cal M}_{X}^{r}$. The latter can be naturally defined by the following characterizing property: ${\bm x}\in {\cal M}_{X}^{r}$ if for some $\beta_{0}({\bm x})$ the operator $\hat{H}_{{\bm\lambda}}({\bm x};\beta)$ has a non-degenerate ground state for all ${\bm\lambda}$ and all $\beta>\beta_{0}$. The subspace ${\cal M}_{X}^{r}$ is a very convenient tool for applying the low-temperature limit. In particular, ${\cal L}^{\beta}$ for large $\beta$ forms a family of bundles over ${\cal M}_{X}^{r}\times {\cal T}$. Therefore, they are all equivalent (isomorphic), and  have the same Chern class, i.e., with a minimal abuse of notation we can write $c_{1}({\cal L})$. More formally, given a driving protocol ${\bm s}$, for $\beta> \max(\beta_{0}({\bm s}(\tau)))$, the Chern numbers $c_{1}({\cal L}^{\beta})f_{\alpha}({\bm s})$ are $\beta$-independent. Therefore, there is a well-defined low-temperature limit
\begin{eqnarray}
\label{n-limit} n_{\alpha}({\bm s})\equiv \lim_{\beta\to\infty}n_{\alpha}({\bm s};\beta)=c_{1}({\cal L})f_{\alpha}({\bm s}),
\end{eqnarray}
where the rhs means that $c_{1}({\cal L})f_{\alpha}({\bm s})\equiv c_{1}({\cal L}^{\beta})f_{\alpha}({\bm s})$, calculated for large enough, still finite $\beta$.

Integer quantization in the low-temperature adiabatic limit can be formulated as a stronger statement in terms of the LD function ${\cal S}({\bm Q})$, or equivalently in terms of $z_{{\bm\lambda}}$. It includes three statements. (i) The topological term $\zeta_{{\bm\lambda}}({\bm s};\beta)$ has a well-defined low-temperature $\beta\to\infty$ limit
\begin{eqnarray}
\label{topological-factor-limit} \zeta_{{\bm\lambda}}^{T}({\bm s})=\prod_{\alpha\in X_{1}}\lambda_{\alpha}^{n_{\alpha}({\bm s})}, \;\; n_{\alpha}({\bm s})=c_{1}({\cal L})f_{\alpha}({\bm s}),
\end{eqnarray}
expressed in terms of Chern numbers. This has been demonstrated earlier in this subsection. (ii) The purely adiabatic contribution $\Lambda_{{\bm\lambda}}({\bm s};\beta)$ possesses time-reversal symmetry $\Lambda_{-{\bm\lambda}}({\bm s};\beta)=\Lambda_{{\bm\lambda}}({\bm s};\beta)$, which is inherited from the time-reversal symmetry $\omega_{-{\bm\lambda}}({\bm x};\beta)=\omega_{{\bm\lambda}}({\bm x};\beta)$. The latter property can be validated in various ways. We will restrict ourselves to mentioning that physically this is equivalent to the even character of the LD function of a non-driven system, i.e., when the parameters $({\bm E},{\bm W})$ do not change with time. (iii) The purely geometric correction to the topological term vanishes $\Lambda_{{\bm\lambda}}^{G}({\bm s};\beta)\to 0$ in the low-temperature $\beta\to\infty$ limit, or equivalently
\begin{eqnarray}
\label{geometric-lT-limit} \lim_{\beta\to\infty}\zeta_{{\bm\lambda}}({\bm s};\beta)=\zeta_{{\bm\lambda}}^{T}({\bm s})
\end{eqnarray}

Properties (i)-(iii) result in the following form of the LD function in the low-temperature adiabatic limit
\begin{eqnarray}
\label{S-low-temp-adiab} {\cal S}({\bm Q};\tau_{D},\beta)&\approx& \tau_{D}{\cal S}^{A}\left(\tau_{D}^{-1}({\bm Q}-{\bm n}({\bm s}));\beta\right), \nonumber \\ {\cal S}^{A}\left(-{\bm J};\beta\right)&=&{\cal S}^{A}\left({\bm J};\beta\right)
\end{eqnarray}
where ${\cal S}^{A}\left({\bm Q};\tau_{D},\beta\right)$ is obtained as a Legendre transform of $\Lambda_{{\bm\lambda}}^{A}({\bm s};\beta)$. In particular, Eq.~(\ref{S-low-temp-adiab}) identifies the low-temperature adiabatic average current $\bar{{\bm Q}}({\bm s})={\bm n}({\bm s})$, thus providing a Chern class representation for $\bar{{\bm Q}}({\bm s})$.

\subsection{Low-temperature calculation of the geometric contribution}
\label{zeta-low-temp}

In this subsection, we perform a calculation of the geometric factor $\zeta_{{\bm\lambda}}({\bm s};\beta)$ in the low-temperature limit, i.e., derive the expression given by Eq.~(\ref{geometric-lT-limit}). In the first manuscript of this series, we identified the subspace ${\cal M}_{X}^{r}=\widetilde{{\cal M}}_{X}-D_{X}^{r}$ of robust parameters by explicitly describing the discriminant space $D_{X}^{r}$ of non-robust parameters. In this subsection we accomplish a more moderate task by showing that ${\cal M}_{X}\subset{\cal M}_{X}^{r}$, i.e., that a driving protocol that avoids simultaneous degeneracy of the energy and barrier data (we reiterate that this is how the subspace ${\cal M}_{X}$ is defined) is a robust protocol in the sense of low-temperature behavior, and derive Eq.~(\ref{geometric-lT-limit}) for such protocols. The validity of Eq.~(\ref{geometric-lT-limit}) for the driving protocols that go through ${\cal M}_{X}$ will be instrumental for the explicit identification of the broader space ${\cal M}_{X}^{r}$.

We fix a driving protocol ${\bm s}$, which is restricted to ${\cal M}_{X}$ at all times, and the argument ${\bm\lambda}\in {\cal T}$ of the generating function, and perform a direct calculation of the geometric factor $\zeta_{{\bm\lambda}}({\bm s};\beta)$. By definition, the latter is represented by a factor in the relation
\begin{eqnarray}
\label{zeta-as-factor} |u(1)\rangle=\zeta_{{\bm\lambda}}({\bm s})|u(0)\rangle,
\end{eqnarray}
where the time-dependent distribution $|u(\tau)\rangle$, represented by ${\bm u}(\tau)=(u_{j}(\tau)|j\in X_{0})$, satisfies the conditions
\begin{eqnarray}
\label{u-of-tau} \hat{H}_{{\bm\lambda}}({\bm s}(\tau);\beta)|u(\tau)\rangle&=&\omega_{{\bm\lambda}}({\bm s}(\tau);\beta)|u(\tau)\rangle, \nonumber \\ \langle u(\tau)|\partial_{\tau}u(\tau)\rangle&=&0,
\end{eqnarray}
for all $0\le \tau \le 1$. Note that Eqs.~(\ref{u-of-tau}) represent a system of equations for standard adiabatic theory, with the second condition ensuring proper treatment of the geometric phase (known as the Berry phase for evolution with time-dependent Schr\"odinger equation).

We further partition the (proper) time segment $[0,1]$ into alternating $0$- and $1$-segments $\tilde{I}_{a}=[\tau'_{a-1},\tau_{a}]$ and $I_{a}=[\tau_{a},\tau'_{a}]$, respectively, exactly in the same way as it was done for average currents in previous article of the series. We reiterate that by construction ${\bm s}(\tau_{a}),{\bm s}(\tau'_{a})\in U_{0}$, with $U_{0}$ and $U_{1}$ being the parameter subspaces with no degeneracy in the energy and barrier data, respectively (by definition ${\cal M}_{X}=U_{0}\cup U_{1}$). So we may introduce the nodes $k_{a}$ and $k'_{a}$, with $k'_{a-1}=k_{a}$, that correspond to the minimal values of $E_{i}(\tau_{a})$ and $E_{i}(\tau'_{a})$, respectively. A simple analysis shows that on a $0$-segment $\tilde{I}_{a}$, in the $\beta\to\infty$, limit the ground state of the twisted master operator is the same as in the untwisted case, i.e., it is localized in the lowest-energy node $u_{j}(\tau)=\zeta_{a}(\tau)\delta_{j}^{k_{a}}$, whereas, the second condition in Eq.~(\ref{u-of-tau}), as applied to our case, requires $\zeta_{a}$ to be time-independent. Therefore, we can introduce the sets of non-zero complex factors $\zeta_{a}$ and $\zeta'_{a}$, so that $u_{j}(\tau_{a})=\zeta_{a}\delta_{j}^{k_{a}}$ and $u_{j}(\tau'_{a})=\zeta'_{a}\delta_{j}^{k'_{a}}$, respectively. Note that $\zeta'_{a-1}=\zeta_{a}$. Then, in the low-temperature limit the geometric contribution is represented by the product
\begin{eqnarray}
\label{geometric-seg-prod} \zeta_{{\bm\lambda}}({\bm s};\beta)=\prod_{a}(\zeta'_{a}/\zeta_{a}),
\end{eqnarray}
and the only thing we need to do is to identify the ratios $\zeta'_{a}/\zeta_{a}$.

In the low-temperature limit, the ground state $|u(\tau)\rangle$ on a $1$-segment can be found asymptotically exactly by replacing the original graph with the relevant minimal spanning tree $\tilde{X}_{{\bm W}_{a}}$. Physically, we are neglecting the transition over the links with the highest values of the barriers $W_{\alpha}$, which can be done, as long as the graph stays connected. The ground-state problem on a tree can be solved exactly, since twisting can be eliminated by a gauge transformation, while the ground state for the untwisted case is represented by the Boltzmann distribution. Since evolution with a standard (untwisted) master operator conserves the total population, the second condition in Eq.~(\ref{u-of-tau}) boils down to the requirement for the gauge-transformed ground state to be represented by the normalized Boltzmann distribution with some time-independent multiplicative factor, i.e., in the original representation on a $1$-segment $I_{a}$ we have
\begin{eqnarray}
\label{u-of-tau-1-segment} u_{j}(\tau)=\zeta_{a}h_{j}\rho_{j}^{B}({\bm E}(\tau);\beta),
\end{eqnarray}
where ${\bm h}=(h_{j}|j\in X_{1})$ represents the gauge transformation that eliminates twisting. The latter requirement means that for any link $\alpha$ on the spanning tree with $\partial\alpha=\{i,j\}$, we have $\lambda_{i\alpha}=\lambda_{j\alpha}^{-1}=h_{i}^{-1}h_{j}$. It is convenient to choose $h_{k_{a}}=1$, so that the set of above constraints has a unique solution: $h_{j}$ is given by the product of $\lambda_{\alpha}$ over all links ${\alpha}$ that belong to the path $l_{k_{a}j}(\tilde{X}_{{\bm W}_{a}})$ that connects the reference node $k_{a}$ to node $j$ through the minimal spanning tree. Recalling the definition of the integer-valued current ${\bm Q}_{ij}(\tilde{X}_{{\bm W}})$, associated with the path $l_{ij}(\tilde{X}_{{\bm W}})$, followed by substituting the obtained value of ${\bm h}$ into Eq.~(\ref{u-of-tau-1-segment}) we arrive at
\begin{eqnarray}
\label{u-of-tau-1-segment-2} u_{j}(\tau)=\zeta_{a}\rho_{j}^{B}({\bm E}(\tau);\beta)\prod_{\alpha\in X_{1}}\lambda_{\alpha}^{Q_{k_{a}j,\alpha}(\tilde{X}_{{\bm W}_{a}})},
\end{eqnarray}
which, due to $\rho_{j}^{B}({\bm E}(\tau_{a});\beta)\to\delta_{j}^{k_{a}}$ and $\rho_{j}^{B}({\bm E}(\tau'_{a});\beta)\to\delta_{j}^{k'_{a}}$ for $\beta\to\infty$, results in
\begin{eqnarray}
\label{ratio-of-zeta} \zeta'_{a}/\zeta_{a}=\prod_{\alpha\in X_{1}}\lambda_{\alpha}^{Q_{k_{a}k'_{a},\alpha}(\tilde{X}_{{\bm W}_{a}})}.
\end{eqnarray}
Substituting the obtained expression into
Eq.~(\ref{geometric-seg-prod}) and recalling the expression for the
average low-temperature adiabatic current $\bar{Q}({\bm s})$,
generated by a good pumping protocol, we arrive at
\begin{eqnarray}
\label{geometric-lT-limit-derived} \lim_{\beta\to\infty}\zeta_{{\bm\lambda}}({\bm s};\beta)=\prod_{\alpha\in X_{1}}\lambda_{\alpha}^{\bar{Q}_{\alpha}({\bm s})}.
\end{eqnarray}

The obtained relation [Eq.~(\ref{geometric-lT-limit-derived})] demonstrates the vanishing $\lim_{\beta\to\infty}\Lambda_{{\bm\lambda}}^{G}({\bm s};\beta)=0$ of the purely geometric correction in the low-temperature limit, or equivalently the statement of Eq.~(\ref{geometric-lT-limit}), and also identifies \begin{eqnarray}
\label{current-equals-chern} \bar{{\bm Q}}({\bm s})={\bm n}({\bm s})
\end{eqnarray}
the low-temperature adiabatic integer-valued average current as the Chern numbers, associated with the corresponding $2$-dimensional cycles ${\bm f}({\bm s})$.

\section{Chern-class representations for quantized currents}
\label{chern-II}

This section is focused on some aspects of the Chern class representation of $\bar{{\bm Q}}({\bm s})$. In subsection~\ref{integer-Chern-revisit} we revisit the integer case and make an explicit connection ${\cal M}_{X}^{r}=\widetilde{{\cal M}}_{X}-D_{X}^{r}$ between the extended subspace ${\cal M}_{X}^{r}$, introduced in the context of the generating function in subsection~\ref{generating-Chern-low-temp} as the limiting space of the temperature-dependent non-degenerate ground-state spaces ${\cal M}_{X}^{r}(\beta)$ with the reduced discriminant set $D_{X}^{r}$ of the non-robust parameters, introduced in the previous manuscript of the series for a more general rational case. In subsection~\ref{Q-rational-Chern} we discuss a Chern class representation for the rational case.

\subsection{Chern class representation for the integer case revisited}
\label{integer-Chern-revisit}

In subsection~\ref{generating-Chern-low-temp} we introduced the space ${\cal M}_{X}^{r}\supset {\cal M}_{X}$ that extends the subspace ${\cal M}_{X}\subset \widetilde{{\cal M}}_{X}$ of good parameters. The extended space ${\cal M}_{X}^{r}$ has been identified as a result of studying the low-temperature adiabatic limit of the generating function for the full counting statistics in the integer-quantization case, i.e., the pdf of the generated current, more precisely the LD function ${\cal S}({\bm Q})$. In the first manuscript of the series, while considering a more general rational case, we extended the subspace ${\cal M}_{X}=\widetilde{{\cal M}}_{X}-D_{X}$ by withdrawing certain unnecessary cells of $D_{X}$ which resulted in a reduced discriminant set $D_{X}^{r}\subset D_{X}$. In this subsection we demonstrate for the integer case that ${\cal M}_{X}^{r}=\widetilde{{\cal M}}_{X}-D_{X}^{r}$.

To show that ${\cal M}_{X}^{r}\subset \widetilde{{\cal M}}_{X}-D_{X}^{r}$, we use a purely topological argument to demonstrate that if a point ${\bm x}=({\bm E},{\bm W})\in D_{\omega}$, i.e., belongs to a cell $D_{\omega}$, with ${\cal D}_{\omega}\ne 0$, then ${\bm x}$ cannot belong to ${\cal M}_{X}^{r}$. We reiterate that, according to the winding-index picture of quantized currents, introduced in section~VIII of the first manuscript of the series for a more general case of rational quantization, further applied to the particular case of integer quantization, the discriminant space $D_{X}$ is represented by a union of the closed cells $\bar{D}_{\omega}$, the latter obtained applying the topological closure to the corresponding open cells $D_{\omega}$, labeled by $\omega=(jj',\alpha\alpha',p)$, where $j\ne j'$ and $\alpha\ne \alpha'$ is a pair of distinct nodes and links, respectively, whereas $p$ is an ordering $\alpha_{1}<\ldots< \alpha_{r}< \{\alpha,\alpha'\}< \alpha_{r+1}<\ldots <\alpha_{|X_{1}|}$ of the graph links. The cell $D_{\omega}$ consists of all parameter sets $(\bm{E},\bm{W})$ with the constraints $E_{j}=E_{j'}< E_{i}$ for any $i\ne j,j'$, and $W_{\alpha_{1}}<\ldots< W_{\alpha_{r}}< W_{\alpha}=W_{\alpha'}< W_{\alpha_{r+1}}<\ldots <W_{\alpha_{|X_{1}|}}$. Stated equivalently the cell $D_{\omega}$ describes the parameter sets with simultaneous lowest-energy degeneracy $E_{j}=E_{j'}$ and barrier degeneracy $W_{\alpha}=W_{\alpha'}$, whereas we have a specific ordering of the barrier heights, given naturally $W_{\alpha}=W_{\alpha'}$. We also reiterate that ${\cal D}_{\omega}=\bar{\bm{Q}}(\bm{s})$ represents the low-temperature adiabatic current generated over a driving protocol $\bm{s}$ that winds around a point $\bm{x}\in D_{\omega}$ in its small neighborhood without intersecting with the cell $D_{\omega}$. As demonstrated in section~VIII of the first manuscript of the series, the current $\bar{\bm{Q}}(\bm{s})$ does not depend on a particular choice of a point $\bm{x}\in D_{\omega}$, as well as a driving protocol $\bm{s}$, as long as the latter belongs to a small enough neighborhood of $\bm{x}$ and winds around $D_{\omega}$ once. The described property reflects the topological nature of the low-temperature adiabatic limit of the generated current.

To show that ${\cal M}_{X}^{r}\subset \widetilde{{\cal M}}_{X}-D_{X}^{r}$ we use the proof by contradiction concept. Suppose not, i.e., ${\bm x}\in {\cal M}_{X}^{r}$. Since ${\cal M}_{X}^{r}$ is open (i.e., also contains some neighborhood of ${\bm x}$), and since the cells $D_{\omega}$ have codimension $2$, we can consider a $2$-dimensional disc ${\cal A}$ centered in ${\bm x}$ that is contained in the aforementioned neighborhood and intersects $D_{\omega}$ transversally. Let ${\bm s}$ be a driving protocol that goes over its boundary once. Then ${\bm s}$ is restricted to ${\cal M}_{X}^{r}$, and ${\cal A}$ is spanned on ${\bm s}$, as well as ${\cal A}\subset{\cal M}_{X}^{r}$. By the winding index representation the generated over the driving protocol $\bm{s}$ low-temperature adiabatic current $\bar{{\bm Q}}({\bm s})={\cal A}\cdot {\cal D}_{X}={\cal D}_{\omega}\ne 0$ is non-zero.

On the other hand, by the Chern class representation we have $\bar{{\bm Q}}({\bm s})=c_{1}({\cal L}){\bm f}({\bm s})$. We further reiterate that $\bm{f}(\bm{s})$ is represented by its components $f_{\alpha}(\bm{s})$, associated with the graph links $\alpha$, as described in some detail in subsection~\ref{generating-Chern-adiabatic}, that map a torus $T^{2}\cong S^{1}\times S^{1}$ to ${\cal M}_{X}^{r}\times {\cal T}$, and referred to as $2$-dimensional cycles in ${\cal M}_{X}^{r}\times {\cal T}$. The disc ${\cal A}$, spanned on $\bm{s}$ can be viewed as a map from a standard disc $D^{2}=\{(x,y)\in \mathbb{R}^{2}| x^{2}+y^{2}\le 1\}$ to ${\cal M}_{X}^{r}$ that extends the driving protocol $\bm{s}$ viewed as a map from a circle $S^{1}$ that represents the disc boundary to the space ${\cal M}_{X}^{r}$ of robust parameters. That delivers an extension of the map $f_{\alpha}(\bm{s}):S^{1}\times S^{1}\to {\cal M}_{X}^{r}\times {\cal T}$ to the map $\tilde{f}_{\alpha}({\cal A}):D^{2}\times S^{1}\to {\cal M}_{X}^{r}\times {\cal T}$ that maps $D^{2}\times S^{1}$ to ${\cal M}_{X}^{r}\times {\cal T}$. Therefore the cycle $f_{\alpha}(\bm{s})$ represents the boundary of $\tilde{f}_{\alpha}({\cal A})$ [which can be formally stated as $f_{\alpha}(\bm{s})=\partial \tilde{f}_{\alpha}({\cal A})$, where $\partial$ has a meaning of a boundary operator], i.e., $f_{\alpha}(\bm{s})$ can be viewed as the restriction of the map $\tilde{f}_{\alpha}({\cal A})$ from its domain $D^{2}\times S^{1}$ to the domain's boundary $S^{1}\times S^{1}$. The value of the Chern class $c_{1}({\cal L})$ evaluated at a boundary is always zero; this intrinsic property of the Chern classes, which reflects their topological (homological) character, is demonstrated for our particular case explicitly in appendix~\ref{Chern-class}. This implies $\bar{{\bm Q}}({\bm s})=c_{1}({\cal L}){\bm f}({\bm s})=c_{1}({\cal L})\partial\tilde{\bm{f}}({\cal A})=0$, which contradicts $\bar{{\bm Q}}({\bm s})={\cal D}_{\omega}\ne 0$, obtained earlier.


We are now in a position to show that $\tilde{{\cal M}}_{X}-D_{X}^{r}\subset {\cal M}_{X}^{r}$. Since in subsection~\ref{zeta-low-temp} we have established ${\cal M}_{X}\subset {\cal M}_{X}^{r}$, it is sufficient to demonstrate that ${\bm x}\in D_{\omega}$ with ${\cal D}_{\omega}=0$ implies ${\bm x}\in {\cal M}_{X}^{r}$. To achieve that, we build explicitly an asymptotically exact, in the $\beta\to\infty$ limit, approximation for the non-degenerate ground state $|u_{{\bm\lambda}}({\bm y};\beta)\rangle$ of the twisted master operator $\hat{H}_{{\bm\lambda}}({\bm y};\beta)$ for all ${\bm\lambda}\in{\cal T}$ and all ${\bm y}$ in some small neighborhood of ${\bm x}$. In the integer case under consideration our cell is labeled $\omega=(jj',\alpha\alpha',p)$ by a pair of distinct nodes, a pair of distinct links and an ordering $p$ of the set of links, as described in some detail earlier in this subsection. In particular in the cell, i.e. for $({\bm E},{\bm W})\in D_{\omega}$, we have $E_{j}=E_{j'}$ and $W_{\alpha}=W_{\alpha'}$. Let $\tilde{X}_{\omega}$ and $\tilde{X}'_{\omega}$ be the maximal spanning trees that correspond to the degeneracy resolutions $W_{\alpha}<W_{\alpha'}$ and $W_{\alpha'}<W_{\alpha}$, respectively. Then ${\cal D}_{\omega}={\bm Q}_{jj'}(\tilde{X}_{\omega})-{\bm Q}_{jj'}(\tilde{X}'_{\omega})$, and therefore ${\cal D}_{\omega}=0$ implies $l_{jj'}(\tilde{X}_{\omega})=l_{jj'}(\tilde{X}'_{\omega})$, so that both can be denoted by $l_{jj'}(\omega)$. Since the ground state population in the neighborhood of ${\bm x}$ is substantially restricted to the the lowest-energy nodes $j$ and $j'$, the existence of a unique preferred path $l_{jj'}(\omega)$ allows the gauge transformation trick, developed in subsection~\ref{zeta-low-temp}, to study the generating function to be applied. Namely we approximate the ground state of the twisted master operator on the original graph with its counterpart associated with the subgraph represented by the path $l_{jj'}(\omega)$. Since the aforementioned subgraph is a tree the ground state of the associated twisted master operator is represented by a gauge transformed Boltzmann distribution, as described in subsection~\ref{zeta-low-temp}, which in the low-temperature limit is concentrated in the nodes $j$ and $j'$. This results in the following explicit expression for the low-temperature limit for the (non-degenerate) twisted ground state
\begin{eqnarray}
\label{u-irrelevant-degeneracy} |u_{{\bm\lambda}}({\bm E},{\bm W};\beta)\rangle&=&Z_{\omega}^{-1}(\beta)e^{-\beta E_{j}}|j\rangle \nonumber \\ &+&Z_{\omega}^{-1}(\beta)e^{-\beta E_{j}'}\prod_{\alpha\in l_{jj'}(\omega)}\lambda_{\alpha}|j'\rangle \nonumber \\ Z_{\omega}(\beta)&\equiv&e^{-\beta E_{j}}+e^{-\beta E_{j'}},
\end{eqnarray}
and in the low-temperature limit Eq.~(\ref{u-irrelevant-degeneracy}) reproduces the expression for the twisted ground state obtained for $({\bm E},{\bm W})\in {\cal M}_{X}$, i.e., no simultaneous degeneracy, in subsection~\ref{zeta-low-temp}. Finally, we emphasize that we were able to obtain the expression of Eq.~(\ref{u-irrelevant-degeneracy}) due to the existence of a unique preferred path $l_{jj'}(\omega)$ in some neighborhood of ${\bm x}$, the latter property resulting from ${\cal D}_{\omega}=0$, and such a construction would not work for ${\cal D}_{\omega}\ne 0$.

\subsection{Chern class representation of rational quantized currents}
\label{Q-rational-Chern}

In this subsection we present a Chern class representation for the pumped current $\bar{{\bm Q}}({\bm s})$ for the rational quantization case. Compared to the integer case, the discussed representation for the rational case is much more involved. There are two reasons for that. First, Chern classes are integer-valued by nature, and to be able to reproduce rational numbers some additional framework needs to be introduced. Second, in the integer case the Chern class representation results from a strong property of the generating function, established in subsection~\ref{zeta-low-temp}: the geometric contribution to the generating function in the low-temperature adiabatic case becomes purely topological, and has a very simple form (as we established explicitly) in terms of the relevant Chern numbers. We analyzed some simple examples with permanent degeneracy, and despite the existence of explicit expressions for average currents, the expressions even for the low-temperature adiabatic limit of the generating functions look complicated. The rational Chern class formula for $\bar{{\bm Q}}({\bm s})$, presented in this subsection, should be viewed as just a translation of the averaging formula to the language of fiber bundles and their characteristic classes. Since in the rational case the discussed representation is more involved and less understood, we will restrict ourselves to formulating the rational Chern class representation and presenting some arguments in its support, postponing the details to some other publication focusing on the topological aspects of the stochastic pumping effect.

An ability to obtain rational numbers is provided by bringing in $K$-theory with rational coefficients \cite{Atiyah}. Conventional topological $K$-theory that can be also referred to as $K$-theory with integer coefficients deals with vector bundles, i.e., the bundles whose fibers are represented by finite-dimensional vector spaces (linear bundles that play an important role in our Chern class picture of stochastic currents represent a particular case of vector bundles of rank $1$). Elements of $K(X)$, associated with a topological space $Y$ are vector bundles, or strictly their equivalence classes. First of all, isomorphic bundles represent the same element of $K(Y)$. There is a coarser equivalence relation also imposed, referred to as {\it stable equivalence}, where we say that two bundles $E$ and $E'$ are stable equivalent $E\sim E'$ if there is a bundle $F$ so that $E\oplus F\cong E'\oplus F$, i.e., the bundles become isomorphic after being summed with some bundle $F$. Note that we can form their direct sum $E\oplus E'$ and tensor product $E\otimes E'$ by applying the operation of a direct sum and tensor product of vector spaces fiberwise. To convert $K(Y)$ to an abelian group, or more precisely a ring we apply the Grothendieck group construction by adding to $K(Y)$ ``negative bundles'', so that an element of $K(Y)$ is represented by a formal difference $E- E'$ with the stable equivalence relation. Equivalence classes $[E]-[E']$ that represent the elements of $K(X)$ are often referred to as stable bundles. Note that the construction that involves ``negative'' bundles is analogous to the one used to obtain integer numbers $\mathbb{Z}$ out of natural ones $\mathbb{N}$.

Before starting a brief description of the rational $K$-theory we consider a simple example that shades light on the importance of the stable equivalence  condition. Let $\mathbb{T}(S^{2})$ be the tangent bundle over a $2$-dimensional sphere, whose fiber $\mathbb{T}_{x}(S^{2})$ consists of all tangent vectors to the sphere that originate at point $x\in S^{2}$, where the sphere is assumed to be embedded $S^{2}\subset \mathbb{R}^{3}$ into the $3$-dimensional space in a standard way. It is intuitive that the tangent bundle $\mathbb{T}_{x}(S^{2})$ is {\it non-trivial}, i.e., its fibers may not be parallelized globally (i.e., for all points of the sphere) in a continuous fashion. The normal bundle ${\cal N}(S^{2})$, whose fibers are represented by the vectors orthogonal to the sphere, is trivial though, since it allows a global continuous non-zero {\rm section}, i.e., associating in a continuous fashion with any point in the sphere a non-zero vector in the fiber, represented, e.g., by a unit normal vector directed outside of the sphere. The direct sum $\mathbb{T}(S^{2})\oplus {\cal N}(S^{2})=S^{2}\times \mathbb{R}^{3}$ is, however, trivial, since the total space of the bundle is given by a cartesian product of the base $S^{2}$ with the fiber $\mathbb{R}^{3}$, with the latter represented by the $3$-dimensional space the sphere is embedded into. Therefore, the non-trivial tangent bundle $\mathbb{T}(S^{2})$ is stable equivalent to a trivial bundle. On a more general note in the stable equivalence world one can identify the normal bundle as ${\cal N}(Y)=-\mathbb{T}(Y)$ the negative tangent bundle for any compact topological space $Y$, and up to a trivial bundle.

$K$-theory with rational coefficients associates with any space $Y$ an algebra $K(Y;\mathbb{Q})$ over rational numbers, i.e., elements of $K(Y;\mathbb{Q})$ can be added, multiplied, and multiplied by rational numbers. Elements of $K(Y;\mathbb{Q})$ are represented by formal superpositions of complex vector bundles over $Y$ with rational coefficients. Similar to the integer case, the
algebra $K(Y;\mathbb{Q})$ is given by imposing the stable equivalence relation on the set of representatives, which looks absolutely the same as in the integer case. Addition and multiplication are naturally defined by the operations of direct sum and tensor product of vector bundles. Elements of $K(Y;\mathbb{Q})$ are referred to as rational stable bundles.

Stable equivalence combined with rational coefficients allows complete analysis of vector bundles, which means rational stable bundles in terms of their Chern classes. The Chern classes are powerful topological invariants of vector bundles over complex numbers (i.e., the fibers are vector spaces over $\mathbb{C}$). However, they generally cannot distinguish between any two non-equivalent bundles. Stable equivalence together with rational coefficients turns the situation around, namely a rational stable bundle is fully characterized by the set of its Chern classes.

When characterizing the rational stable bundles in terms of their Chern classes, it is more natural to deal with the {\it Chern characters}, rather than Chern classes themselves. The Chern character $ch({\cal L})$ of ${\cal L}\in K(Y;\mathbb{Q})$ is a universal combination of Chern classes of ${\cal L}$, and is represented by a rational superposition of even-dimensional cocycles in $Y$, i.e., linear functionals on cycles of $Y$. Stated  equivalently a $2k$-cocycle can be evaluated on a $2k$-cycle of $Y$, resulting in a rational number. Note that for example the first Chern class $c_{1}$ is a $2$-cocycle with integer coefficients. The convenience of Chern characters is determined by their algebraic properties $ch({\cal L}\oplus{\cal L}')=ch({\cal L})+ch({\cal L}')$ and $ch({\cal L}\otimes{\cal L}')=ch({\cal L})\cdot ch({\cal L}')$. For a linear (i.e., $1$-dimensional) bundle ${\cal L}$ we have
\begin{eqnarray}
\label{ch-linear-bundle} ch(L)=e^{c_{1}({\cal L})}=1+c_{1}({\cal L})+\frac{1}{2!}(c_{1}({\cal L}))^{2}+\ldots,
\end{eqnarray}
where the the term $(c_{1}({\cal L}))^{2}$ and higher ones should be evaluated on the cycles of dimension $4$ and higher, respectively.

The reason why we brought in $K$-theory with rational coefficients is that if instead of having one preferred line bundle ${\cal L}$ over ${\cal M}_{X}\times{\cal T}$ that describes the pumped current $\bar{{\bm Q}}({\bm s})$ via its first Chern class $c_{1}({\cal L})$, as is in the integer case, we have a set of line bundles, and we can average them with rational weights, resulting in an ``averaged bundle'' ${\cal L}\in K({\cal M}_{X}\times{\cal T};\mathbb{Q})$ and further express the pumped current via its Chern character $ch({\cal L})$.

Recalling the definition of $2$-cycles $f_{\alpha}({\bm s})$ in the space ${\cal M}_{X}\times{\cal T}$, labeled by the links $\alpha\in X_{1}$ of our network $X$, the Chern class representation of $\bar{{\bm Q}}({\bm s})$ has the form
\begin{eqnarray}
\label{Q-chern-character} \bar{Q}_{\alpha}({\bm s})=ch({\cal L})f_{\alpha}({\bm s}), \;\; {\cal L}\equiv \frac{1}{n_{X}}\sum_{({\bm k},\tilde{{\bm X}})}{\cal L}_{{\bm k},\tilde{{\bm X}}},
\end{eqnarray}
where ${\cal L}\in K({\cal M}_{X}\times {\cal T};\mathbb{Q})$, and a construction of the line bundle ${\cal L}_{{\bm k},\tilde{{\bm X}}}$, associated with a degeneracy resolution, labeled by $({\bm k},\tilde{{\bm X}})$ is presented in appendix~\ref{construct-L-stable}.

The Chern class representation [Eq.~(\ref{Q-chern-character})] can be justified by recasting it in an equivalent form
\begin{eqnarray}
\label{Q-chern-character-2} \bar{Q}_{\alpha}({\bm s})=\frac{1}{n_{X}}\sum_{({\bm k},\tilde{{\bm X}})}c_{1}\left({\cal L}_{{\bm k},\tilde{{\bm X}}}\right)f_{\alpha}({\bm s}),
\end{eqnarray}
followed by applying the definitions and the averaging formula.

The extension of the Chern class representation [Eq.~(\ref{Q-chern-character})] to the extended subspace ${\cal M}_{X}^{r}\supset {\cal M}_{X}$ is as follows. The above inclusion generates an algebra homomorphism $K({\cal M}_{X}^{r}\times{\cal T};\mathbb{Q})\to K({\cal M}_{X}\times{\cal T};\mathbb{Q})$ that boils down to restriction of a bundle to a subspace. The space ${\cal M}_{X}^{r}$ satisfies the following properties: (i) the rational stable bundle ${\cal L}\in K({\cal M}_{X}\times{\cal T};\mathbb{Q})$ allows a unique extension to a stable bundle ${\cal L}^{r}\in K({\cal M}_{X}^{r}\times{\cal T};\mathbb{Q})$, and (ii) ${\cal L}^{r}$ cannot be extended to a rational stable bundle over $U\times {\cal T}$ for any open $U\supset {\cal M}_{X}^{r}$ and $U\ne {\cal M}_{X}^{r}$. The space ${\cal M}_{X}^{r}$ is fully characterized by above properties, and so is ${\cal L}^{r}$. For a driving protocol ${\bm s}$ that stays entirely in ${\cal M}_{X}^{r}$ the Chern class representation reads
\begin{eqnarray}
\label{Q-chern-character-3} \bar{Q}_{\alpha}({\bm s})=ch({\cal L}^{r})f_{\alpha}({\bm s}).
\end{eqnarray}

It is also instructive to note that since the line bundles ${\cal L}_{{\bm k},\tilde{{\bm X}}}$ are trivial over $U_{{\bm E}}$ and $U_{{\bm W}}$, so is there direct sum $n_{X}{\cal L}=\bigoplus_{({\bm k},\tilde{{\bm X}})}{\cal L}_{{\bm k},\tilde{{\bm X}}}$. This means that $n_{X}{\cal L}$ can be viewed as a complex vector bundle of dimension $n_{X}$ obtained by gluing together the trivial bundles defined over $U_{{\bm E}}$ and $U_{{\bm W}}$, so that the Chern class representation for the current can be recast in a form
\begin{eqnarray}
\label{Q-chern-character-4} \bar{Q}_{\alpha}({\bm s})=\frac{1}{n_{X}}c_{1}(n_{X}{\cal L})f_{\alpha}({\bm s}).
\end{eqnarray}

\section{Discussion}
\label{discussion-II}

We have demonstrated that topological invariants, such as Chern classes/numbers, naturally appear in classical stochastic processes when we consider counting statistics of generated currents under the condition of periodic time-dependence of kinetic rates. We determined that the integer quantization of currents, observed in the adiabatic low temperature limit, is a special manifestation of such Chern numbers. The regime of fractional quantization remains much less understood. We calculated the counting statistics in this regime for a simple model with 2 nodes and $N$ links. The result appears reminiscent the averaging formula, in which a weighted sum of the reference integer circulating (conserved) currents ${\bm Q}$, associated with legitimate degeneracy resolutions, as described in detail in the first manuscript of the series, is replaced by the weighted sum of the $e^{i{\bm\chi}{\bm Q}}$ terms with the same weights. We leave the question of possible generalization to the case of arbitrary graphs as an open problem.

We also demonstrated that the space of robust parameters for cyclic protocols that generate the integer-quantized currents, which are robust with respect to small perturbations of the driving protocol, is the same as the space of parameters that allow the standard adiabatic approach to be applied to the generating function calculation for all values of its argument. This results in the identification of a topological invariant (based on first Chern class) of the counting statistics that is directly related to the quantized average current. The described relation provides a simple approach to classifying topological properties of the counting statistics.

Our results are only a first step towards understanding the topologically protected behavior of the counting statistics. We focused on the low-temperature adiabatic $\bar{{\bm Q}}({\bm s})\equiv\lim_{\beta\to\infty}\lim_{\tau_{D}\to\infty}{\bar {\bm Q}}({\bm s};\tau_{D};\beta)$ limit. Nevertheless, we showed that topological invariants in counting statistics can be defined for arbitrary temperatures and
frequencies. In this more general situation they no longer directly lead to the quantization of average currents, but still may correspond to some other effects.

Other types of  topological invariants in counting statistics, which have not been discussed in the present article, are possible. For example, in \cite{ren-10prl} fractional quantization of currents in a model without detailed balance was found. When detailed balance is not in place the twisted master operator is no longer Hermitian. The corresponding geometric phases may have different topological properties due to existence of the so-called exceptional points \cite{sinitsyn-09review}. Other topological invariants in counting statistics, called $\mathbb{Z}_{2}$-invariants, have been recently found and studied \cite{sinitsyn-10jstat}.

We would also like to note that a direct relation between the twisted master operators and Hermitian
Hamiltonians suggests that all of the phenomena we have discussed in this series in terms of applications to stochastic processes can be reinterpreted in terms of the corresponding properties of quantum mechanical systems. For example, in the present manuscript we have provided a simple and intuitive procedure for calculating integer topological invariants of the operator eigenstates phase space, parametrized by the periodic variables ${\bm\lambda}$ and ${\bm s}$. In quantum mechanical applications, such topological
invariants may describe Chern classes of vector bundles, associated with Bloch bands. Hence, the simple
classification approach, based on the averaging formula, may appear useful for studying topological properties of some Bloch bands.

In both manuscripts of the series we have been studying generation of currents for stochastic motion in networks, or in other words on graphs, i.e., in the discrete setting. Similar questions can be asked for continuous setting, i.e, in the case of Langevin stochastic processes in continuous spaces, e.g., manifolds, where the role of the master equation/operator is played by its Fokker-Planck counterpart. When the configuration space $X$ where stochastic dynamics occurs has non-contractible $1$-dimensional cycles (loops) global stochastic currents (fluxes) can be introduced (see, e.g., \cite{ccmt09}) and observed in experiment. The detailed balance condition corresponds to potential forces that describe effects of advection, and driving can be described by time dependence of a potential function $V({\bm x})$. A generic potential function is of Morse type, i.e., its critical points are isolated and have non-degenerate matrices of second derivatives. In the low-temperature $\beta\to\infty$ limit, the Boltzmann distribution $\rho^{B}({\bm x})=Z^{-1}e^{-\beta V({\bm x})}$ is concentrated in a small neighborhood (the lower the temperature, the smaller the neighborhood) of the global minimum of the potential $V({\bm x})$. There are two mechanisms that generate current
in the continuous case. In the first case, consider a situation when the potential evolves in time in a way that the positions ${\bm x}_{j}$ of the local minima stay the same, while the local minimal values $V({\bm x}_{j})$ of the potential function are subject to changes in time. When the global minimum at, say ${\bm x}_{i}$, loses  its global character, delegating it, say to ${\bm x}_{k}$, a current will be produced due to the probability flux from a neighborhood of ${\bm x}_{i}$ to the neighborhood of ${\bm x}_{k}$, with the flux concentrated around the minimal energy path, the latter generally passing between a set of some other local minima via elementary paths that go through the transition states, i.e., critical points ${\bm y}_{\alpha}$ with a single unstable mode. The described process can be readily analyzed by applying the approaches, developed in this series, to a network, whose nodes and links are represented by ${\bm x}_{j}$ and ${\bm y}_{\alpha}$, respectively, while the rates are computed using standard transition rate theory for overdamped Langevin dynamics (see, e.g., \cite{htb90}). The second mechanism for generating current is just simple periodic motion of a global minimum. Strictly speaking, current generation is a combination of the above two mechanisms.

Most of the structures of our present study can be extended to the continuous case using more technical, rather than conceptual effort. In particular, a space of good parameters that has a property of describing generic potential functions can be relatively easily introduced. The question of identifying the robust parameter space, in the sense of our present consideration, is more complex. Current quantization for Langevin processes in continuous spaces, which is currently under study, will be presented in our further publications.

Finally, the fact that certain properties of fluctuations in mesoscopic systems can be robustly controlled by periodic driving protocols should be of interest for characterization and control of mesoscopic structures, such as electronic nanoscale conductors \cite{makhlin-mirlin} and molecular motors \cite{astumian-11rev,sinitsyn-09jcp}.

\appendix

\section{First Chern class and relevant Chern numbers}
\label{Chern-class}
\begin{figure}
\centerline{\includegraphics[width=2.0in]{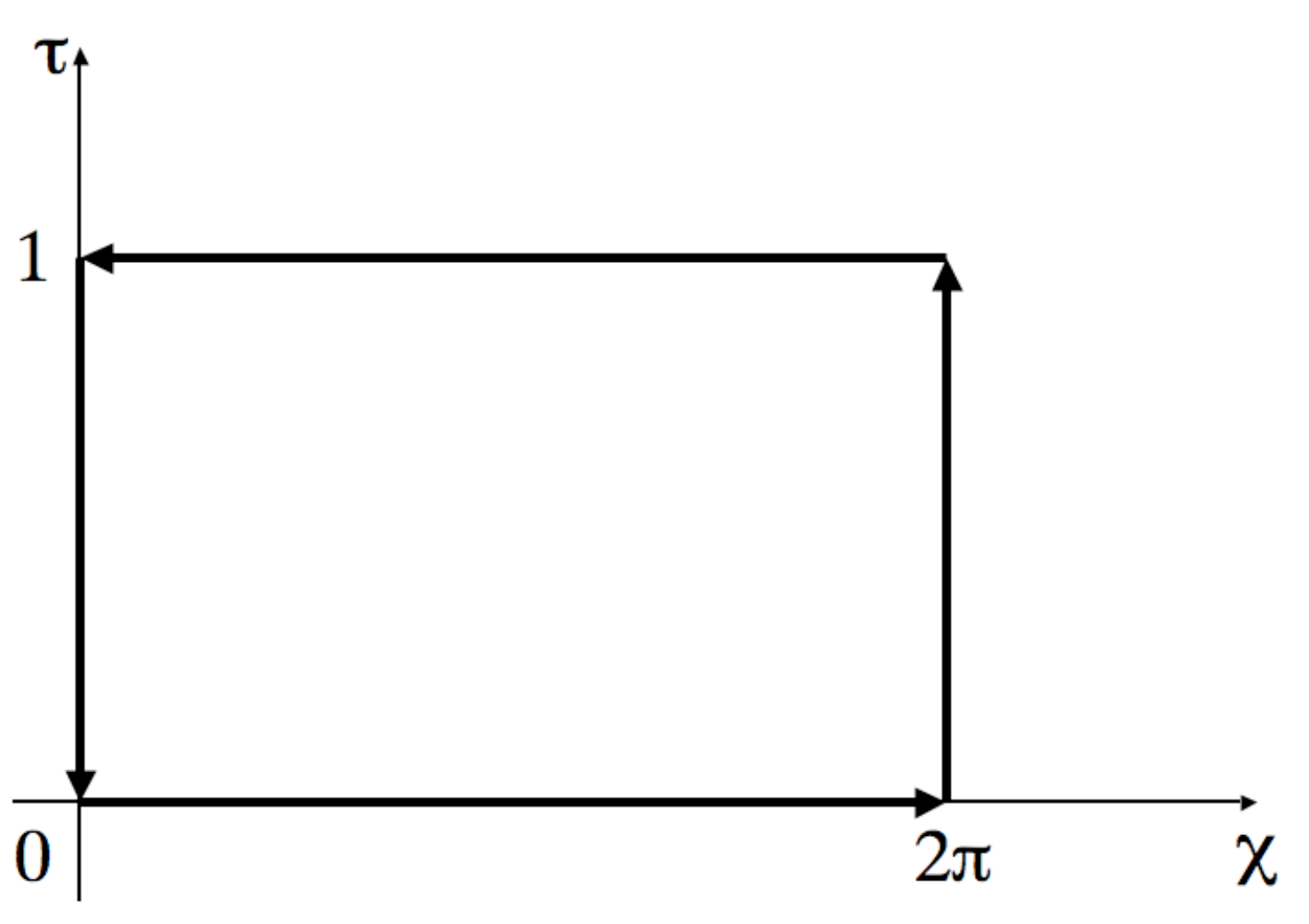}}
  \caption{ Contour of integration}
\label{2level-split}
\end{figure}
In this appendix we present a simple and self-contained description of the first Chern class and provide all necessary information for our applications. Chern classes $c_{j}$ with $j=1,2,\ldots$ are useful
invariants of complex vector  bundles. In our applications we will be dealing with linear, i.e., dimension one bundles, which are fully characterized by their first Chern class $c_{1}$.

When computing a generating function, we deal with a family of twisted
master operators, $\hat{H}_{{\bm\lambda}}({\bm x})$, where ${\bm\lambda}\in {\cal T}$ is the argument of the generating function, whereas ${\bm x}$ represents a set of parameters, e.g., the node energies
$E_{j}$ and the barriers $W_{\alpha}$, i.e., ${\bm x}=({\bm E},{\bm W})$ with ${\bm E}=(E_{j}|j\in X_{0})$ and ${\bm W}=(W_{\alpha}|\alpha\in X_{1})$.
Note that the operator $\hat{H}_{{\bm\chi}}({\bm x})$ becomes Hermitian upon the transformation, $\hat{H}_{{\bm\lambda}}({\bm x})\rightarrow e^{\beta \hat{E}/2} \hat{H}_{{\bm\lambda}}({\bm x})e^{-\beta \hat{E}/2}$, however, its components are generally not real. This  means that eigenvalues of $\hat{H}_{{\bm\lambda}}({\bm x})$ are real, whereas the eigenvectors have generally complex components. An adiabatic calculation of the generating function for a given driving protocol is straightforward, as long as the ground state of $\hat{H}_{{\bm\lambda}}({\bm x})$ is non-degenerate for all ${\bm\lambda}\in {\cal T}$ and all values ${\bm x}$ of parameters involved in the driving protocol \cite{sinitsyn-07epl,sinitsyn-09review}. We reiterate that we refer to such parameters sets as non-degenerate ground-state parameters. The (temperature-dependent) space of such parameters ${\bm x}=({\bm E},{\bm W})$ has been denoted by ${\cal M}_{X}^{r}(\beta)$. Actually, one of the main results of this manuscript is the demonstration that in the low-temperature adiabatic limit the average current is quantized and is robust with respect to perturbations of the driving protocol as long as the latter is restricted to the subspace, ${\cal M}_{X}^{r}\subset \widetilde{{\cal M}}_{X}$, of robust parameters, which is a kind of a low-temperature limit space for ${\cal M}_{X}^{r}(\beta)$; the precise construction is presented in subsection~\ref{generating-Chern-low-temp}.

Having restricted ourselves to the subspace of non-degenerate ground-state parameters, we note that with each point ${\bm y}=({\bm x},{\bm\lambda})=({\bm E},{\bm W};{\bm\lambda})$ we can associate a complex line (complex $1$-dimensional vector space) of eigenstates $|u\rangle$, such that
$\hat{H}_{{\bm\lambda}}({\bm x})|u\rangle=\omega_{{\bm y}}|u\rangle$, where $\omega_{{\bm y}}$ is the largest eigenstate of $\hat{H}_{{\bm\lambda}}({\bm x})$. Considering the described above eigenstates for all $\hat{H}_{{\bm\lambda}}({\bm x})$ we obtain a space ${\cal L}$, referred to as the {\it total space} of a {\it vector bundle}, that consists of points $({\bm x},{\bm\lambda},|u\rangle)$. The space of eigenstates, $|u\rangle$, can be viewed as {\it fibers} of the vector bundle, and are parametrized by $({\bm x},{\bm\chi})$.

We are now in a position to introduce the first Chern class $c_{1}({\cal L})$ of ${\cal L}$. We start with noting that the fiber ${\cal L}_{{\bm y}}$ over ${\bm y}=({\bm x},{\bm\lambda})$ consists of the states $|u\rangle$ that differ by just normalization factors, and for any ${\bm y}$ we can locally, i.e., in some neighborhood of ${\bm y}$, fix the normalization factors in a smooth way, i.e., build a smooth function $|u_{{\bm\lambda}}({\bm x})\rangle$ of $({\bm x},{\bm\lambda})$ with $\hat{H}_{{\bm\lambda}}({\bm x})|u_{{\bm\lambda}}({\bm x})\rangle=\omega_{{\bm y}}|u_{{\bm\lambda}}({\bm x})\rangle$. Such a function is referred to as a local {\it section} of ${\cal L}$. We further introduce a vector-potential
\begin{eqnarray}
\label{define-connection} A_{j}({\bm y})=\frac{\langle u({\bm y})|\partial_{j}u({\bm y})\rangle}{\langle u({\bm y})|u({\bm y})\rangle}, \;\;\; \partial_{j}=\partial/\partial y_{j},
\end{eqnarray}
where we introduce the left eigenstates, defined by  $\langle u \vert \hat{H}_{{\bm\lambda}}({\bm x})=\langle u \vert \omega_{{\bm y}}$, and the scalar product is defined as the sum of the product of corresponding components of the left and right eigenstates. We define the {\it  curvature of the vector-potential} by
\begin{eqnarray}
\label{curvature} F_{ij}({\bm y})=\partial_{i}A_{j}({\bm y})-\partial_{j}A_{i}({\bm y}),
\end{eqnarray}
the latter satisfying a property
\begin{eqnarray}
\label{curvature-closed} \partial_{i}F_{jk}({\bm y})+\partial_{k}F_{ij}({\bm y})+\partial_{j}F_{ki}({\bm y})=0.
\end{eqnarray}
Eq. (\ref{curvature-closed})  is usually referred to as the property
of $F_{ij}({\bm y})$ to be {\it closed}. It is very important to note that the vector-potential depends on the particular choice of a local section, i.e. for $|u'({\bm y})\rangle=e^{\varphi({\bm y})}|u({\bm y})\rangle$ we have
\begin{eqnarray}
\label{gauge-transform} A'_{j}({\bm y})=A_{j}({\bm y})+\partial_{j}\varphi({\bm y}),
\end{eqnarray}
however, Eq.~(\ref{gauge-transform}) represents a gauge transformation, so that the curvature $F_{ij}({\bm y})$ is not sensitive to the local section choice, and there fore $F_{ij}({\bm y})$ is defined globally and depends only on ${\cal L}$ and the scalar product.

Being interested in  applications that involve integration of the curvatures over $2$-dimensional cycles we define that  two closed forms, $F_{ij}({\bm y})$ and $F'_{ij}({\bm y})$, are {\it equivalent} if $F'_{ij}({\bm y})=F_{ij}({\bm y})+\partial_{i}a_{j}({\bm y})-\partial_{j}a_{i}({\bm y})$ for some globally defined smooth $a_{i}({\bm y})$, and hereafter use the notation $[F]$ for the equivalence class of $F_{ij}({\bm y})$. The equivalence class $[F]$ does not depend on a particular choice of the scalar product, so that the first Chern class, defined by $c_{1}({\cal L})\equiv [F]$, depends on ${\cal L}$ only. We will also demonstrate that the first Chern class is the same for equivalent bundles ${\cal L}'\simeq {\cal L}$, e.g., when the fibers of ${\cal L}'$ can be continuously deformed to the fibers of ${\cal L}$, so that the first Chern class $c_{1}({\cal L})$ is a {\it topological invariant} of ${\cal L}$.

We conclude this brief description of the first Chern class of line bundles with showing how to evaluate the Chern class on $2$-dimensional cycles in the parameter space represented by points ${\bm y}=({\bm x},{\bm\lambda})$, resulting in {\it Chern numbers}, followed by demonstrating that all Chern numbers are represented by integers. By analogy with ``standard'' $1$-dimensional cycles, we define a $2$-dimensional cycle as a compact oriented $2$-dimensional manifold mapped in a smooth way to the parameter space of ${\bm y}$. We will restrict ourselves to the case when the domain manifold is a $2$-dimensional torus $T^{2}$ with coordinates $0 \le \tau \le 1$ and $0 \le \chi \le 2\pi$, also denoted by $(\xi^{1},\xi^{2})$, so that a cycle $f$ is represented by a smooth doubly periodic map ${\bm y}({\bm\xi})$. The Chern number $c_{1}({\cal L})f$, which can be viewed as a result of evaluation of the Chern class on the cycle $f$, is given by
\begin{eqnarray}
\label{Chern-number} c_{1}({\cal L})f=\int_{0}^{1}d\tau\int_{0}^{2\pi}\frac{d\chi}{2\pi i}\epsilon^{ab}\frac{\partial y^{i}}{\partial \xi^{a}}\frac{\partial y^{j}}{\partial \xi^{b}}F_{ij}({\bm y}({\bm\xi})),
\end{eqnarray}
where $\epsilon^{12}=-\epsilon^{21}=1$ is a $2$-dimensional
antisymmetric object, and hereafter we assume summation over repeating
indices. For our applications, $\tau$ will represent the dimensionless
time on the driving protocol and $\chi$ is a
phase of some component of a multi-variant argument of the generating function.

To establish the integer nature of Chern numbers we {\it pull-back} the bundle ${\cal L}$ to $T^{2}$ along $f$ that results in a bundle over $T^{2}$, denoted $f^{*}{\cal L}$, whose fiber $f^{*}{\cal L}_{{\bm\xi}}$ is ${\cal L}_{{\bm y}({\bm\xi})}$. Following the general procedure described earlier [Eqs.~(\ref{define-connection}) and (\ref{curvature})] we construct a gauge field $\bar{A}_{a}({\bm\xi})$ and curvature $\bar{F}_{ab}({\bm\xi})$. Due to covariance we have $\bar{A}_{a}({\bm\xi})=(\partial y^{j}/\partial\xi_{a})A_{j}({\bm y}({\bm\xi}))$ and $\bar{F}_{ab}({\bm\xi})=(\partial y^{i}/\partial\xi_{a})(\partial y^{j}/\partial\xi_{b})F_{ij}({\bm y}({\bm\xi}))$, so that
\begin{eqnarray}
\label{Chern-number-2} c_{1}({\cal L})f=\int_{0}^{1}d\tau\int_{0}^{2\pi}\frac{d\chi}{2\pi i}\epsilon^{ab}\bar{F}_{ab}({\bm\xi}).
\end{eqnarray}
We further view a torus $T^{2}$ as a rectangle $[0,1]\times [0,2\pi]$ with the opposite sides being identified, and build a section of $f^{*}{\cal L}$ over the rectangle by fixing the normalization of $|u(0,\chi)\rangle$ in some arbitrary way (which is obviously always possible to do), followed by extending the normalization to the whole rectangle by imposing the condition $\langle u(\tau,\chi)|\partial_{\tau}u(\tau,\xi)\rangle=0$. This results in a section $|u({\bm\xi})\rangle$, which is quasi-periodic, i.e., $|u(\tau,\chi+2\pi)\rangle=|u(\tau,\chi)\rangle$ and $|u(\tau+1,\chi)\rangle=\zeta(\chi)|u(\tau,\chi)\rangle$, with $\zeta(\chi+2\pi)=\zeta(\chi)$. Due to the quasi-periodic nature of $|u({\bm\xi})\rangle$ the vector potentials on the opposite sides of the rectangle differ by just gauge transformations [Eq.~(\ref{gauge-transform})]:
\begin{eqnarray}
\label{A-opposite} \bar{A}_{\tau}(\tau,2\pi)&=&\bar{A}_{\tau}(\tau,0), \nonumber \\ \bar{A}_{\chi}(1,\chi)&=&\bar{A}_{\chi}(0,\chi)+\partial_{\chi}\zeta(\chi)/\zeta(\chi)
\end{eqnarray}
Integrating the vector potential over the border of the rectangle in the counterclockwise direction we arrive at
\begin{eqnarray}
\label{integral-border} \frac{1}{2\pi i}\oint_{C}d\xi^{j}\bar{A}_{j}({\bm\xi})=\int_{0}^{2\pi}\frac{d\chi}{2\pi i\zeta(\chi)}\frac{\partial\zeta(\chi)}{\partial\chi}=n[\zeta(\chi)],
\end{eqnarray}
where $n[\zeta(\chi)]$ is the number of times $z\eta\in \mathbb{C}$ winds over the zero point in the complex plane $\mathbb{C}$ while $\chi$ varies from $0$ to $2\pi$. Since Stokes' theorem  (see e.g., \cite{GP}) identifies the l.h.s. of Eq.~(\ref{integral-border}) with the r.h.s. of Eq.~(\ref{Chern-number-2}), we can express the Chern number as
\begin{eqnarray}
\label{Chern-number-3} c_{1}({\cal L})f=n[\zeta(\chi)].
\end{eqnarray}
Where the r.h.s. of the later is the winding index. We wish to emphasize that, from its construction, $\zeta(\chi)$ may be interpreted as the adiabatic geometric factor associated with a cycle of adiabatic dynamics with a twisted master operator $\hat{H}_{\lambda}$, where $\lambda=e^{i\chi}$. Therefore, $\zeta(\chi)$ describes the dependence of the geometric factor on the argument of the generating function, whereas the Chern number $c_{1}({\cal L})f$ represents an integer topological invariant of $\zeta(\chi)$, and Eq.~(\ref{Chern-number-3}) will play a key role in identifying the Chern numbers as the components of the quantized currents in the low-temperature adiabatic regime of stochastic pumping.

In subsection~\ref{integer-Chern-revisit} we have used the intrinsic property of Chern classes to vanish being evaluated at boundaries. Here we demonstrate this property explicitly for our particular application. For a map ${\cal A}:D^{2}\to {\cal M}_{X}^{r}$, where $D^{2}$ is a standard $2$-dimensional disc defined in subsection~\ref{integer-Chern-revisit}, we have a map $\tilde{f}_{\alpha}:D^{2}\times S^{1}\to {\cal M}_{X}^{r}\times {\cal T}$, defined by $\tilde {f}_{\alpha}({\cal A}(\eta,\chi))=({\cal A}(\eta),\bm{\lambda}^{\alpha}(\chi))$, where the components of $\bm{\lambda}^{\alpha}(\chi)$ are: $\lambda_{\alpha}^{\alpha}(\chi)=e^{i\chi}$, and $\lambda_{\gamma}^{\alpha}(\chi)=1$ for $\gamma\ne \alpha$, as described at the end subsection~\ref{generating-Chern-adiabatic} in the contest of defining the set $\bm{f}$ of $2$-dimensional cycles. The map $\tilde{f}_{\alpha}({\cal A})$ is defined in a way so that the cycle $f_{\alpha}(\bm{s})=\partial \tilde{f}_{\alpha}({\cal A})$ represents its boundary, i.e., $f_{\alpha}$ is obtained from $\tilde{f}_{\alpha}({\cal A})$ by restricting the domain $D^{2}\times S^{1}$ to its boundary $\partial (D^{2}\times S^{1})=S^{1}\times S^{1}$. We further apply the expression of Eq.~(\ref{Chern-number-2})
\begin{eqnarray}
\label{Chern-number-4} c_{1}({\cal L})f_{\alpha}(\bm{s})=\int_{S^{1}\times S^{1}}d^{2}\xi\epsilon^{ab}\bar{F}_{ab}({\bm\xi}),
\end{eqnarray}
written in terms of some coordinates of the torus. We further note that the curvature is actually defined over $D^{2}\times S^{1}$ by $\bar{F}_{ab}(\bm{\zeta})=(\partial \tilde{f}_{\alpha}^{i}/\partial\zeta_{a})(\partial \tilde{f}_{\alpha}^{j}/\partial\zeta_{b})F_{ij}({\bm y}({\bm\xi}))$, and being restricted to its boundary $S^{1}\times S^{1}$ reproduces the curvature in Eq.~(\ref{Chern-number-4}). This allows the general (multi-dimensional) Stokes formula to be applied
\begin{eqnarray}
\label{Stokes-formula} \int_{S^{1}\times S^{1}}\bar{F}=\int_{D^{2}\times S^{1}}dF,
\end{eqnarray}
since $S^{1}\times S^{1}=\partial(D^{2}\times S^{1})$. The Stokes formula, presented in Eq.~(\ref{Stokes-formula}) using the language of differential forms 
(\cite{GP} \cite{WW}) has the following coordinate representation
\begin{eqnarray}
\label{Stokes-formula-2} \int_{S^{1}\times S^{1}}d^{2}\xi\epsilon^{ab}\bar{F}_{ab}({\bm\xi}) \nonumber \\ =\int_{D^{2}\times S^{1}}d^{3}\zeta\epsilon^{abc}\partial_{a}\bar{F}_{bc}({\bm\zeta}),
\end{eqnarray}
where $\epsilon^{abc}$ is the $3$-dimensional antisymmetric object. Due to the definition of the curvature [Eq.~(\ref{curvature})] the curvature is closed, i.e., $dF=0$, or in coordinates is given by Eq.~(\ref{curvature-closed}). Therefore the integrand in the r.h.s. of Eq.~(\ref{Stokes-formula-2}) turns to zero, which combined with Eq.~(\ref{Chern-number-4}) yields $c_{1}({\cal L})\partial\tilde{f}_{\alpha}({\cal A})=0$, i.e., vanishing of the Chern class evaluated at a boundary cycle.

\section{Construction of the current generating stable bundle}
\label{construct-L-stable}

According to Eq.~(\ref{Q-chern-character}) the current generating stable bundle ${\cal L}$ is obtained via averaging of ${\cal L}_{{\bm k},\tilde{{\bm X}}}$ over all allowed degeneracy resolutions. In this appendix we present a construction of the line bundle ${\cal L}_{{\bm k},\tilde{{\bm X}}}$, associated with a degeneracy resolution, labeled by $({\bm k},\tilde{{\bm X}})$. The construction is based on gluing together a set of bundles defined over the open sets that cover the bundle base ${\cal M}_{X}\times {\cal T}$.

The subspace ${\cal M}_{X}$ is obviously covered by a family of open sets
\begin{eqnarray}
\label{M-X-cover-family} {\cal M}_{X}=\bigcup_{[{\bm E}]\in {\cal X}_{0}}U_{[{\bm E}]} \;\;\cup \bigcup_{[{\bm W}]\in {\cal O}({\cal X}_{1})}U_{[{\bm W}]}\, ,
\end{eqnarray}
where $U_{[{\bm E}]}$ and $U_{[{\bm W}]}$ consist of all $({\bm E}',{\bm W}')$ so that $[{\bm E}]'=[{\bm E}]$ and $[{\bm W}]'=[{\bm W}]$, respectively. Note that $U_{[{\bm E}]}\cap U_{[{\bm E}']}=\emptyset$ and $U_{[{\bm W}]}\cap U_{[{\bm W}']}=\emptyset$ for $[{\bm E}]\ne [{\bm E}]'$ and $[{\bm W}]\ne [{\bm W}]'$, respectively. In particular there are no triple or higher-order intersections among the sets of the family. Given a degeneracy resolution $({\bm k},\tilde{{\bm X}})$, we can build complex line bundles ${\cal L}_{{\bm k}}^{[{\bm E}]}$ and ${\cal L}_{\tilde{{\bm X}}}^{[{\bm W}]}$ over $U_{[{\bm E}]}\times {\cal T}$ and $U_{[{\bm W}]}\times {\cal T}$, respectively. The fiber of ${\cal L}_{{\bm k}}^{[{\bm E}]}$ over $({\bm E},{\bm W},{\bm\lambda})\in U_{[{\bm E}]}\times {\cal T}$ is spanned on the state $|k_{{\bm E}}\rangle$, whereas the fiber of ${\cal L}_{\tilde{{\bm X}}}^{[{\bm W}]}$ over $({\bm x},{\bm\lambda})=({\bm E},{\bm W},{\bm\lambda})\in U_{[{\bm W}]}\times {\cal T}$ is represented by the eigenstates $|u\rangle$ of the operator $\hat{H}_{{\bm\lambda}}({\bm x};\beta)|_{\tilde{X}_{{\bm W}}}$ with the lowest eigenvalue. The operator $\hat{H}_{{\bm\lambda}}({\bm x};\beta)|_{\tilde{X}_{{\bm W}}}$ is obtained by restricting the twisted master operator $\hat{H}_{{\bm\lambda}}({\bm x};\beta)$ to the spanning tree $\tilde{X}_{{\bm W}}$, i.e., the rates associated with the withdrawn links are set to zero. The fibers of ${\cal L}_{{\bm k}}^{[{\bm E}]}$ and ${\cal L}_{\tilde{{\bm X}}}^{[{\bm W}]}$ over all points $({\bm x},{\bm\lambda})\in (U_{[{\bm E}]}\cap U_{[{\bm W}]})\times {\cal T}$ should be glued together. This is done by just projecting an eigenstate $|u\rangle$ onto $|k_{{\bm E}}\rangle$. Since there are no triple intersections of the open subsets, no consistency conditions have to be taken care of.

\end{document}